\documentclass[aps,prd,reprint,onecolumn,notitlepage,nofootinbib,superscriptaddress]{revtex4-1}
\usepackage{lipsum}
\usepackage{amsmath}
\usepackage{graphicx}
\usepackage{color}
\usepackage{tikz}
\usepackage{verbatim}
\usepackage{hyperref}
\usepackage{amssymb}
\usepackage{bm}
\usepackage{makecell}
\usepackage{booktabs}
\usepackage[justification=raggedright]{caption}
\usepackage{subcaption}
\usepackage{multirow}
\graphicspath{ {./}}
\usepackage{tabularx}
\newcolumntype{Y}{>{\centering\arraybackslash}X}

\newcommand{\rom}[1]{\textup{\uppercase\expandafter{\romannumeral#1}}}

\usepackage{float}
\begin{document}

\title{Phantom Menace in general Palatini $f(R,\phi)$ theories}

\author{Rahul Thakur\footnote{rahul19@iiserb.ac.in}}
\affiliation{Department of Physics, Indian Institute of Science Education and Research Bhopal, Madhya Pradesh - 462066, India}
\author{Abhijith Ajith\footnote{abhijith.ajith.421997@gmail.com}}
\affiliation{Department of Physics, Indian Institute of Science Education and Research Bhopal, Madhya Pradesh - 462066, India}
\author{Sukanta Panda\footnote{sukanta@iiserb.ac.in}}
\affiliation{Department of Physics, Indian Institute of Science Education and Research Bhopal, Madhya Pradesh - 462066, India}
\author{Archit Vidyarthi\footnote{architmedes@gmail.com}}
\affiliation{Department of Physics, Indian Institute of Science Education and Research Bhopal, Madhya Pradesh - 462066, India}
\affiliation{Department of Physics, Indian Institute of Technology Bombay, Powai, Mumbai, Maharashtra - 400076, India}

\begin{abstract}
We study general $f(R,\phi)$ theories in Palatini formalism and attempt to constrain the behavior of ones that could support both inflationary and late-time expansion era in a unified model. In particular, we find conditions for which the theories remain consistent in weak gravity regimes as well as cosmic expansion eras in both early and late universe. Assuming that the curvature part of the $f(R,\phi)$ behaves as Starobinsky gravity, we assess post-inflation dynamical stability of the theory in Einstein frame and proceed to isolate two distinct fixed points that provide a stable late-time accelerating universe. Comparison with DESI, Cosmic Chronometers, and SNeIa datasets adds more stringent constraints to the behavior of the theory near the present epoch, giving us one stable fixed point where expansion is driven by a phantom scalar field. However, time scales of the two fixed points suggest that this fixed point may be transient and may eventually evolve toward a stable expansion stage driven potential domination in the distant future of the universe.
\end{abstract}

\maketitle 

\section{INTRODUCTION}\label{sec:intro}
The universe is currently experiencing an epoch of accelerated expansion, a phenomenon that has been documented over the past century through various observational efforts. Among the recent key sources of evidence are the measurements of luminosity distance and redshift of Type Ia supernovae \cite{riess1998observational, perlmutter1999measurements, Riess:2006fw}, which have played a central role in this discovery. Other important observations, such as the cosmic microwave background radiation \cite{gawiser2000cosmic}, baryon acoustic oscillations \cite{eisenstein2005detection, percival2007measuring}, and the determination of the Hubble constant \cite{riess2009redetermination}, have further contributed to our understanding of this accelerated phase. Due to sheer lack of any information about what's driving this late-time expansion, cosmologists have attributed it to the mysterious \textit{dark energy} which exerts a large negative pressure to counteract gravitational collapse.

Although many theoretical models have been proposed, the true nature of dark energy is still unknown. One of the first and most widely studied models is the $\Lambda$CDM model, which includes the cosmological constant ($\Lambda$) and cold dark matter. This model has been popular, especially among the particle physics community.
However, the severe fine-tuning associated with the observed value of $\Lambda$, along with the coincidence problem \cite{Weinberg1989,Padmanabhan2003}, suggests that a dynamical mechanism may be responsible for cosmic acceleration. This has driven interest in exploring dynamical dark energy models based on scalar fields, where a scalar field is responsible for generating negative pressure, leading to the accelerated expansion of the universe at late times. Depending on the form of the Lagrangian, there are mainly two types of scalar field models studied so far: the quintessence models \cite{zlatev1999quintessence, peccei1987adjusting, ford1987cosmological, peebles2003cosmological, nishioka1992inflation, ferreira1997structure, ferreira1998cosmology, caldwell1998cosmological, carroll1998quintessence, copeland1998exponential, hebecker2000quintessential, Hebecker_2001,Patil:2023rqy,Samanta:2025oqz,Zhang:2025dwu,Li:2025vuh} and the $k$-essence models \cite{armendariz1999k, garriga1999perturbations, armendariz2001essentials, armendariz2000dynamical, chimento2004power, chimento2004extended, scherrer2004purely, chiba2000kinetically, bose2009k}.

Another promising direction involves modifying the gravitational sector itself. Among such theories, $f(R)$ gravity, which generalizes the Einstein-Hilbert action by promoting the Ricci scalar $R$ to a function $f(R)$, has attracted considerable interest~\cite{capozziello2002curvature, capozziello2003curvature, carroll2004cosmic, nojiri2004modified, dunsby2010lambda, faraoni2008f, sotiriou2010, defelice2010}. These models can naturally drive late-time acceleration without introducing exotic matter components. However, pure $f(R)$ models have their limitations. In particular, they often face challenges in fitting both cosmological and local gravitational tests simultaneously, and they may lack the flexibility needed to reproduce a wide range of cosmological behaviors \cite{DeFelice:2010aj}.

To overcome these shortcomings, an extended class of theories has been proposed in which a scalar field $\phi$ is explicitly coupled to the curvature, leading to models of the form $f(R, \phi)$ \cite{hwang1996unified, bahamonde2015generalized, panda2023constant, farajollahi2011cosmic, odintsov2019unification}. These models merge the advantages of both scalar field dynamics and modified gravity, offering a richer framework to explore dark energy and late-time acceleration. More recently, certain classes of $f(R,\phi)$ inflation models have also been shown to produce a slightly enhanced scalar spectral index corresponding to the recent results from Atacama Cosmology Telescope (ACT). The latest ACT data release has disfavored long-standing inflationary paradigms such as Higgs inflation and Starobinsky inflation, though there have been efforts to show that modified dynamics may be able to restore these models to within the acceptable range \cite{Zharov:2025zjg,Gialamas:2025kef,Drees:2025ngb,Antoniadis:2025pfa,Gialamas:2025ofz,Oikonomou:2025htz,Odintsov:2025eiv,Wolf:2025ecy,Ajith:2025rvf}. More importantly, it has been highlighted by some authors that the non-minimal coupling $\phi R$ is well-suited to produce the required enhancement in the scalar spectral index.

An important subtlety in studying $f(R,\phi)$ models lies in the choice of variational principle. In the metric formalism, the action is varied with respect to the metric, and the connection is taken to be Levi Civita. In contrast, the Palatini formalism treats the metric and affine connection as independent variables \cite{sotiriou2010f,Olmo2011,Panda:2022esd}. The resulting equations of motion are second order in derivatives, thereby avoiding the instabilities that typically emerge in higher derivative metric gravity theories. Furthermore, the Palatini approach often leads to new scalar tensor equivalences that are not present in the metric formulation, making it a compelling framework for exploring modified gravity.

A standard technique for revealing the effective degrees of freedom is the application of a Weyl transformation which maps the Jordan frame action into its equivalent Einstein frame counterpart (wherein gravitational sector resembles GR and the modifications are absorbed into scalar field terms with non standard kinetic and potential structures) \cite{faraoni1998conformal,flanagan2004conformal}. In the case of $f(R,\phi)$ models in Palatini formalism, the transformed action exhibits a particularly novel structure: the scalar sector acquires a higher order kinetic interaction term and an effective potential \cite{das2021inflation}. The appearance of higher order kinetic terms places the model within the broad class of $k$-essence theories \cite{armendariz2001essentials,chiba2002tracking}. These models are known to produce rich cosmological behavior, including dynamical dark energy with varying equation of state, attractor solutions, and even phantom like regimes without introducing ghosts \cite{tsujikawa2013quintessence}. 

In this paper, we investigate the cosmological behavior of a class of Palatini $f(R, \phi)$ gravity models in the Einstein frame and examine their viability as an explanation for the observed acceleration of the universe. Our goal is to determine whether such models can provide an improved or complementary description to $\Lambda$CDM while remaining consistent with current observational data. We perform a dynamical system analysis of cosmological evolution which provides a systematic way to classify possible asymptotic states of the Universe by converting the cosmological equations into an autonomous system of differential equations \cite{copeland1998exponential,bahamonde2018dynamical}. Our main goal is to identify conditions under which such models admit stable accelerated attractors at late times and thus providing a framework for dynamical dark energy. In addition, we highlight the role of the induced quartic kinetic interaction term in driving acceleration.  

The paper is organized as follows: In Section \ref{sec:jordan}, we present the theoretical framework for Jordan frame $f(R,\phi)$ gravity theories such that they could effectively describe a late-time accelerated epoch. In Section \ref{sec:einstein-metric}, we review the corresponding Einstein frame action in metric formalism and provide a generalized canonicalization approach without fixing specific forms of potentials and coupling functions. Next, in Section \ref{sec:einstein-palatini}, we briefly look at the Einstein frame action in Palatini formalism and choose a generalized scalar-Starobinsky model (based on consistency conditions). Its dynamical stability is analyzed in Section \ref{sec:dynamical} where we list fixed points that could provide an efficient candidate for dark energy at late-times. Constraints and other bounds for all the relevant parameters are found with respect to late-time observational data in Section \ref{sec:data-analysis}. Finally, Section~\ref{sec:discussion} summarizes our conclusions and outlines possible avenues for future research.  
\section{Jordan Frame $f(R,\phi)$ theories: Early and Late-time Acceleration} \label{sec:jordan}
\noindent We consider the following action in the Jordan frame,
\begin{equation}\label{eq:jordan-action}
     S=\int d^{4}x\sqrt{-g}\left[\frac{1}{2\kappa}f(R,\phi)-\frac{1}{2}\nabla{_\mu}\phi\nabla{^\mu}\phi-V(\phi)\right].
\end{equation}
Here $g$ is the metric determinant, $R$ is the Ricci scalar for our given spacetime, and $\kappa=M_{\rm Pl}^{-2}$ where $M_{\rm Pl}$ is the reduced Planck mass. Deriving the corresponding Einstein field equations and the equation of motion for the scalar field $\phi$, we obtain,
\begin{align}
&FR_{\mu\nu}-\frac{1}{2}fg_{\mu\nu}-\nabla_{\mu}\nabla_{\nu}F+g_{\mu\nu}\square F=\kappa\left[(\nabla_{\mu}\phi\nabla_{\nu}\phi-\frac{1}{2}g_{\mu\nu}\nabla^{\alpha}\phi\nabla_{\alpha}\phi)-Vg_{\mu\nu}\right],\label{eq:jordanfe}\\
&\square\phi+\frac{1}{2}\left(\frac{f_{,\phi}}{\kappa^2}-2V_{,\phi}\right)=0,\label{eq:jordanphie}
\end{align}
where $F=\frac{\partial f}{\partial R}$. Now, for the background spacetime, we consider a spatially flat FLRW metric given by the line element,  
\begin{equation}\label{eq:flrw-metric}
ds^2=-dt^2+a^2(t)(dx^2+dy^2+dz^2),
\end{equation}
where $a(t)$ denotes the scale factor of expansion. Assuming homogeneity in the background, i.e. $\phi\equiv\phi(t)$, 00 and $ii$ components of the Einstein field equations \eqref{eq:jordanfe} can be expressed respectively as, 
\begin{align}
6F(\dot{H}+H^2)+(2\kappa V-f)-6\dot{F}H+\kappa{\dot\phi}^2&=0,\label{eq:jordanfrphife0}\\
-\kappa{\dot\phi}^2+2F(\dot{H}+H^2)+4FH^2+(2\kappa V-f)-4H\dot{F}-2\ddot{F}&=0.\label{eq:jordanfrphifei}
\end{align}
Further, the equation of motion of $\phi$ given by \eqref{eq:jordanphie} now becomes:
\begin{equation}\label{eq:frphi-equation-phi}
\Ddot{\phi}+3H\dot\phi-\frac{1}{2}\left(\frac{f_{,\phi}}{\kappa}-2V_{,\phi}\right)=0.
\end{equation}
These models have been studied extensively in reference to the inflationary paradigm \cite{farajollahi2011cosmic,Myrzakulov:2015qaa,Mathew:2017lvh,Panda:2022esd,Panda:2022can,Kuralkar:2025hoz}. However, as discussed earlier, the allure of modified gravity and scalar-tensor theories (STTs) lies in their ability to explain a variety of phenomena ranging from inflation to late-time acceleration. To demonstrate this, we shall try to find conditions for a general $f(R,\phi)$ theory of the form \eqref{eq:jordan-action} that could provide a stable dark energy candidate. In general $f(R,\phi)$ theories, both $\phi$ and $F$ are important scalar degrees of freedom (DOF) and either can be responsible for late-time expansion. We can find conditions for $\phi$ to support late-time acceleration by rewriting \eqref{eq:frphi-equation-phi} as,
\begin{equation}
    \Box\phi=V_{,\phi}-\frac{f_{,\phi}}{2\kappa}\equiv\frac{\partial U_{\rm eff}}{\partial\phi},
\end{equation}
Now, we can find the condition of existence of an extremum as,
\begin{align}
    \frac{\partial U_{\rm eff}}{\partial\phi}&=0,\nonumber\\
    f_{,\phi}&=2\kappa V_{,\phi}.
\end{align}
And for this extremum to be a stable minimum, we find,
\begin{align}
    \frac{\partial^2 U_{\rm eff}}{\partial\phi^2}&>0,\nonumber\\
    2\kappa V_{,\phi\phi}&>f_{,\phi\phi}.
\end{align}
It is clear that we can go no further without assuming explicit forms for $f(R,\phi)$ and $V(\phi)$. But as this goes against the foundation of this work, we stop here and shift our focus on the other viable scalar DOF: $F$. The trace equation corresponding to \eqref{eq:jordanfe}, without the assumption of homogeneous background, is given by,
\begin{equation}
3\Box{F}+FR-2f=\kappa T=-\kappa[\partial_{\mu}{\phi}\partial^{\mu}{\phi}+4V(\phi)],
\end{equation}
where $T$ represents trace of the total stress-energy tensor that includes the scalar field $\phi$ and other non-gravitational components (that are no longer decoupled from the system at late times). We can write rewrite the trace equation as \cite{Oikonomou:2025qub},
\begin{equation}
\Box{F}=\frac{1}{3}(\kappa T+2f-FR)\equiv\frac{\partial{V_{\rm eff}}}{\partial{F}}.
\end{equation}
Now, for the existence of a local extremum, we should have,
\begin{align}
\frac{\partial{V_{\rm eff}}}{\partial{F}}=&\frac{1}{3}(\kappa T+2f-FR)=0,\nonumber\\
\implies\ \ \ \ \ R=&\frac{1}{F}(2f+\kappa T).
\end{align}
Assuming a de Sitter expansion at late times, $R\approx 12H^2>0$. Thus we obtain,
\begin{equation}
    2f+\kappa T>0.
\end{equation}
For this extremum to be a local minimum, i.e. for a stable late-time cosmic expansion, we require
\begin{align}\label{eq:condition-stable-latetime-expansion}
&\frac{\partial^2{V_{\rm eff}}}{\partial{F^2}}=\frac{1}{3}\left(2\frac{\partial{f}}{\partial{F}}-R-F\frac{\partial R}{\partial F}\right)>0,\nonumber\\
\implies &2\frac{\partial{f}}{\partial{\phi}}\left(\frac{\partial{F}}{\partial{\phi}}\right)^{-1}+F\left(\frac{\partial{F}}{\partial{R}}\right)^{-1}-R>0,
\end{align}
which is true only when $\frac{\partial{F}}{\partial{\phi}},\frac{\partial{f}}{\partial{R}}\neq0$. Similar condition for $f(R)$ gravity has been derived for a stable inflationary era in \cite{Oikonomou:2025qub}. This condition can be further simplified assuming that $f(R,\phi)$ is factorizable as $f=p(R)q(\phi)$, for some functions $p(R)$ and $q(\phi)$. To do so, we list below a few properties of $f(R,\phi)$ and their consequences for $p(R)$ and $q(\phi)$:
\begin{itemize}
    \item $f(R,\phi)$ has 2 mass dimensions (see  \eqref{eq:jordan-action}). Consequently, we choose $p(R)$ to have 2 mass dimensions while $q(\phi)$ remains dimensionless.
    \item $f(R,\phi)$ must be analytic in both $R$ and $\phi$. This is required because, as $\phi\to0$, $f(R,\phi)$ must become some function of $R$. $R\to0$ must also be a well defined limit in order to explain transitions between various epochs of the universe. In factorizable scenarios, analyticity of $p(R)$ and $q(\phi)$ implies that as $\phi\to0$, $q(\phi)$ returns to some finite value, i.e. there exists a non-zero minimum of $q$.
    \item In weak gravity regimes, $f(R,\phi)$ must reduce to the Einstein-Hilbert action coupled with some function of $\phi$, i.e. $p(R)\to R$.
\end{itemize}
The condition \eqref{eq:condition-stable-latetime-expansion} can then be expressed as,
\begin{equation}\label{eq:condition-stable-latetime-expansion-pq}
   \frac{2p}{p_R} + \frac{p_{R}}{p_{RR}}-R>0
\end{equation}
where the subscript $R$ represents a differentiation w.r.t. $R$. By assuming factorizability of $f$, we arrive at the weaker condition \eqref{eq:condition-stable-latetime-expansion-pq} which is valid as long as $p_R,\, p_{RR}\neq0$. Note that this condition is completely independent of the scalar coupling $q(\phi)$. For $p=R+\alpha R^2$, one can verify that the condition is satisfied for $\alpha>0$ in both weak gravity $\alpha R\gg1$ and strong gravity $\alpha R\ll1$ regimes. Performing cosmological perturbation theory as a next step in this analysis can also help probe whether the DOFs $F$ and $\phi$ support an early-universe inflationary era by deriving observables and comparing them against exiting datasets \cite{DeFelice:2010aj,Mathew:2017lvh}.

\section{Einstein Frame: Metric Formalism} \label{sec:einstein-metric}
While Jordan frame can be quite informative in terms of modifications in the gravity sector, cosmological analyses are typically performed by going to the equivalent Einstein frame via Weyl transformations \cite{Panda:2022esd}. This is done by first introducing to the action \eqref{eq:jordan-action} the term $\frac{F}{2\kappa}(R-\chi^2)$ (where $\chi$ is an auxiliary field) and then performing a Weyl transformation given by $g_{\mu\nu}\rightarrow Fg_{\mu\nu}$ to obtain,
\begin{equation}
  S_E=\int d^{4}x\sqrt{-g}\left[\frac{R}{2\kappa}-\frac{3}{2\kappa F^2}\partial_{\mu}{F}\partial^{\mu}{F}-\frac{1}{2F}\partial{_\mu}\phi\partial{^\mu}\phi-\frac{1}{F^2}\left(\frac{1}{2\kappa}(\chi^2 F-f)+V\right)\right],
\end{equation}
where $F\equiv F(\chi,\phi)$. The corresponding equations of motion are obtained as,
\begin{equation}
\frac{1}{2\kappa}{G_{\mu\nu}}=\frac{3}{\kappa F^2}\partial_{\mu}{F}\partial_{\nu}{F}+\frac{1}{F}\partial{_\mu}\phi\partial{_\nu}\phi-{g_{\mu\nu}}\left[\frac{3}{2\kappa F^2}\partial_{\alpha}{F}\partial^{\alpha}{F}+\frac{1}{2F}\partial_{\alpha}{\phi}\partial^{\alpha}{\phi}+\frac{1}{F^2}\left(\frac{1}{2\kappa}(\chi^2 F-f)+V\right)\right],
\end{equation}
For this tensor equation, the 00 component is given as,
\begin{equation}
\frac{-3}{F^2}\dot{F}^2-\frac{\kappa}{F}\dot{\phi}^2-\frac{2\kappa}{F^2}\left(\frac{1}{2\kappa}(\chi^2 F-f)+V\right)=-3{H^2},
\end{equation}
and $ii$ components are,
\begin{equation}
\frac{3}{F^2}\dot{F}^2+\frac{\kappa}{F}\dot{\phi}^2-\frac{2\kappa}{F^2}\left(\frac{1}{2\kappa}(\chi^2 F-f)+V\right)=-2{\dot{H}}-3{H^2}.
\end{equation}
These equations can be used to perform perturbative and late-time analyses similar to the Jordan frame. But instead of these cumbersome equations, we can instead work with their canonicalized fields which greatly simplify calculations and localize modifications compared to standard STTs to certain terms, leaving others intact. Canonicalization is also a necessary step before proceeding with any quantum field theoretical analysis, such as finding unitarity violation scales or checking renormalizability of the theory \cite{Joshi:2023otx}. Proceeding with canonicalization, we can express,
\begin{equation}
\partial_{\mu}{F}=\frac{\partial{F}}{\partial{\phi}}\partial_{\mu}{\phi}+\frac{\partial{F}}{\partial{\chi}}\partial_{\mu}{\chi},
\end{equation}
which gives,
\begin{align}
\partial_{\mu}{\phi}&=\left(\partial_{\mu}{F}-\frac{\partial{F}}{\partial{\chi}}\partial_{\mu}{\chi}\right)\left(\frac{\partial{F}}{\partial{\phi}}\right)^{-1},\\
\frac{3}{\kappa F^2}\partial_{\mu}{F}\partial^{\mu}{F}&=\frac{3}{\kappa F^2}\left[\left(\frac{\partial{F}}{\partial{\phi}}\right)^2{\partial{_\mu}\phi\partial^{\mu}\phi} +\left(\frac{\partial{F}}{\partial{\chi}}\right)^2{\partial{_\mu}\chi\partial^{\mu}\chi} +2\left( {\frac{\partial{F}}{\partial{\phi}}}\cdot{\frac{\partial{F}}{\partial{\chi}}}\right)\partial{_\mu}\phi\partial^{\mu}\chi\right].
\end{align}
Now, 
\begin{align}
\partial_{\mu}{\ln{F}}&=\frac{F_{\phi}}{F}\partial_{\mu}{\phi}+\frac{F_{\chi}}{F}\partial_{\mu}{\chi},\\
\partial_{\mu}{(\ln{F}})\partial^{\mu}(\ln{F})&=\left(\frac{F_{\phi}}{F}\right)^2\partial{_\mu}\phi\partial^{\mu}\phi+\left(\frac{F_{\chi}}{F}\right)^2\partial{_\mu}\chi\partial^{\mu}\chi+2\left(\frac{F_{\chi}F_{\phi}}{F^2}\right)\partial{_\mu}\phi\partial^{\mu}\chi.
\end{align}
Let $\sqrt{\frac{2}{\kappa}}\ln{F}=\theta$. Then, we have, 
\begin{equation}
\frac{1}{F^2}\partial_{\mu}{\phi}\partial^{\mu}{\phi}=\exp\left({-\sqrt{\frac{\kappa}{3}}}\theta\right)\partial_{\mu}{\phi}\partial^{\mu}{\phi}=\partial_{\mu}{\phi}\partial^{\mu}{\phi}-\sqrt{\frac{\kappa}{3}}\theta\partial_{\mu}{\phi}\partial^{\mu}{\phi}+\frac{\kappa}{6}\theta^2\partial_{\mu}{\phi}\partial^{\mu}{\phi}+...,
\end{equation}
which gives an infinite series of terms that can be truncated based on the scale of $F$ and the higher order kinetic interactions can be used to check renormalizability and unitarity violation scales. But this method leaves the fate of the term $(\chi^2F-f)$ uncertain. The canonicalization problem is better resolved assuming $f(\chi,\phi)$ is factorizable, i.e. $f(\chi,\phi)=p(\chi)q(\phi)$. In that case,
\begin{equation}
\frac{3}{\kappa F^2}\partial_{\mu}{F}\partial^{\mu}{F}=\frac{3}{\kappa} \left[ \left(\frac{q_{\phi}}{q}\right)^2 {\partial{_\mu}\phi\partial^{\mu}\phi} +\left(\frac{p_{\chi^2\chi}}{p_{\chi^2}}\right)^2 {\partial{_\mu}\chi\partial^{\mu}\chi}+2 \left(\frac{p_{\chi^2\chi}}{p_{\chi^2}}\right)\left(\frac{q_{\phi}}{q}\right) {\partial{_\mu}\chi\partial^{\mu}\phi} \right],
\end{equation}
where the subscripts represent variable with respect to which differentiation is being done. Now, considering, 
\begin{equation}
\sqrt{\frac{3}{\kappa}}\frac{p_{\chi^2\chi}}{p_{\chi^2}}\partial_{\mu}{\chi}=\partial_{\mu}{\Omega},
\end{equation}  
or
\begin{equation}\label{eq:p-chi-canonicalization-metric}
\partial_{\mu}\left[{\frac{3}{\kappa}}\ln(p_{\chi^2})\right]=\partial_{\mu}{\Omega}\implies p_{\chi^2}=\exp{\left(\sqrt{\frac{\kappa}{3}} \Omega \right)}.
\end{equation}
Similarly, 
\begin{equation}
\frac{1}{p_{\chi^2} q}\partial_{\mu}{\phi}\partial^{\mu}{\phi}=\exp{\left(-\sqrt{\frac{\kappa}{3}} \Omega \right)}\frac{1}{q}\partial_{\mu}{\phi}\partial^{\mu}{\phi}=\frac{1}{q}\partial_{\mu}{\phi}\partial^{\mu}{\phi}-\sqrt{\frac{\kappa}{3}}\frac{1}{q}\Omega\partial_{\mu}{\phi}\partial^{\mu}{\phi}+\frac{\kappa}{3}\frac{\Omega^2}{q} \partial_{\mu}{\phi}\partial^{\mu}{\phi}.
\end{equation}
Then, the total kinetic term of $\phi$ is,
\begin{equation}
\left(\sqrt{\frac{3}{\kappa}}\frac{q_{\phi}}{q}+\frac{1}{\sqrt{q}}\right)\partial_{\mu}{\phi}=\partial_{\mu}{\theta}.
\end{equation}
Canonicalization is now complete. But the term $(\chi^2F-f)$ still needs to be expressed in terms of canonicalized fields $\Omega$ and $\theta$. In factorized form,
\begin{equation}
\chi^2F-f=(\chi^2 p_{\chi^2}-p)q.
\end{equation}
Now, if $(\chi^2 p_{\chi^2}-p)$ can be expressed as an explicit function of only $p_{\chi^2}$, then using canonicalization transformations, $(\chi^2F-f)$ can be recast as an interaction between $\Omega$ and $\theta$ fields. While this treatment in Einstein frame is straightforward, the situation is a little more complicated when we follow the Palatini formalism because the auxiliary field is non-dynamical, as we shall see in the next section.

\section{Einstein Frame: Palatini Formalism} \label{sec:einstein-palatini}
\noindent Proceeding with the same Weyl transformation of the Jordan frame action \eqref{eq:jordan-action} but in the Palatini formalism, we arrive at the following Einstein frame action:
\begin{equation}\label{eq:palatini-einstein-action}
  S=\int d^{4}x\sqrt{-{g}}\left[\frac{{R}}{2\kappa}-\frac{1}{2F}\partial{_\mu}\phi\partial{^\mu}\phi-\frac{1}{F^2}\left(\frac{1}{2\kappa}(\chi^2 F-f)+V\right)\right].
\end{equation}
It is clear that $\chi$ in this case is a non-dynamical field \cite{DeFelice:2010aj}. Since it is a superfluous DOF in the system, varying the action with respect to $\chi^2$ can help us obtain a constraint equation for the system,
\begin{equation}
\frac{1}{2F^2}\left(\frac{\partial F}{\partial \chi^2}\right) \partial_{\mu}{\phi}\partial^{\mu}{\phi}+\frac{2}{F^3}\left(\frac{\partial F}{\partial \chi^2}\right)\left(\frac{1}{2\kappa}(\chi^2 F-f)+V\right)-\frac{1}{2\kappa F^2}\left(F+\chi^2 \frac{\partial F}{\partial \chi^2} -F\right)=0,
\end{equation}
which on simplification yields,
\begin{equation}\label{eq:palatini-constaint-chi}
\kappa F \partial_{\mu}{\phi}\partial^{\mu}\phi + F \chi^2 -2f +4\kappa V =0.
\end{equation}
This equation can be used to eliminate the non-dynamical DOF $\chi$ from  \eqref{eq:palatini-einstein-action}, which is necessary to proceed with any physical analysis in this system. We also require this constraint in order to canonicalize $\phi$ given in  \eqref{eq:palatini-einstein-action} where the field redefinition would obviously depend of $F(\chi,\phi)$. We will first try to find valid forms of $p$ for which the the constraint equation can be expressed in a form that could eliminate $F(\chi,\phi)$. It would be easier to find conditions for which the non-dynamical DOF can be eliminated from the system if we can rewrite the constraint equation \eqref{eq:palatini-constaint-chi} by factorizing $f(\chi,\phi)$ as $f(\chi,\phi)=p(\chi)q(\phi)$ and rewriting $p\equiv p(R)$ for typographical ease. Now, we have,
\begin{align}
&\kappa p_{R} q  \partial_{\mu}{\phi}\partial^{\mu}{\phi} +\chi^2 p_{R} q -2pq + 4 \kappa V=0,\nonumber\\
\implies &\chi^2 p_{R} - 2p=-\frac{4\kappa V}{q}-\kappa p_{R} \partial_{\mu}{\phi}\partial^{\mu}{\phi}.\label{eq:palatini-constraint-simplified-terms}
\end{align}
Assuming $p(R)=R\ u(R)$ for the same reasons as earlier, we find,
\begin{equation}
  p_{R}=u(R)+R\frac{du}{dR}.
\end{equation}
This equation has the same complimentary solution as in metric formalism. However, the terms we need to manage are slightly different. From the left hand side (LHS) of \eqref{eq:palatini-constraint-simplified-terms}, we have,
\begin{equation}\label{eq:palatini-differential-equation-2}
  R p_{R}-2p=R^2\frac{du}{dR}-uR.
\end{equation}
Since, we have assumed analyticity of $p$, we restrict ourselves only to certain forms of $p$. In theory, many varied forms of this function may exist, each with their signature phenomenological features and high-energy behavior. We also do not consider any forms that may involve weak gravity cut-offs such as \cite{Hu:2007nk,Starobinsky:2007hu}. The two simplest cases we analyze are listed below as examples:
\begin{itemize}
    \item \textbf{Case 1:} When $p_{R}$ is polynomial in $R$ as $p_{R}=a+bR+cR^2+..$, the complete solution has the form,
    \begin{equation}
        u(R)=\frac{m}{R} + Ap_{R} +B.
    \end{equation}
    Substituting in  \eqref{eq:palatini-differential-equation-2}, we get,
    \begin{align}
        &Rp_{R} - 2p = R^2 \left[-\frac{m}{R^2} +A(b+2cR+...) \right]-m-AR(a+bR+cR^2+...)-BR,\\
        \implies &Rp_{R} - 2p=-2m-(aA+B)R+...
    \end{align}
    By substituting the RHS from this expression in \eqref{eq:palatini-constraint-simplified-terms} and solving the resulting polynomial equation in $\chi^2$, we can completely eliminate $\chi^2$ using the constraint equation. As an example, consider $p=R+\alpha R^2$, i.e. a Starobinsky-like gravity with $a=1$, $b=2\alpha$, $A=2$, $B=-1$, and the other constants set to 0. We, then, find,
    \begin{align}
        &\chi^2=\frac{4\kappa V+\kappa q\partial_{\mu}{\phi}\partial^{\mu}{\phi}}{q(1-2\kappa \alpha \partial_{\mu}{\phi}\partial^{\mu}{\phi})},\\
        \implies &p_{R}q=\frac{q+8\kappa \alpha V }{1-2\kappa \alpha \partial_{\mu}{\phi}\partial^{\mu}{\phi}}.
    \end{align}
    For higher-order polynomials $p(R)$, the expressions become extremely cumbersome and may even require additional information about the relative magnitudes of the potentials and kinetic terms to avoid complex solutions of $\chi$. Fixing these relative magnitudes at this stage would defeat the purpose of the dynamical analysis we shall performed in the next section.
    \item \textbf{Case 2:} When $p_R$ is an exponential function of $R$, $p_{R}=ae^{\lambda R}$ for which,
    \begin{equation}
        u=\frac{m}{R} + A e^{\lambda R}.
    \end{equation}
    Following the same procedure as the previous case, we arrive at the following equation:
    \begin{equation}
        -2m+\left[1+ \ln\left(\frac{p_{R}}{a} \right) \right]\frac{A}{a}  \ln \left(\frac{p_{R}}{a} \right) p_{R} + \kappa p_{R} \partial_{\mu}{\phi}\partial^{\mu}{\phi} = -\frac{4\kappa V}{q},
    \end{equation}
    which gives no straightforward solution except the trivial case where $\lambda=0$, i.e. $p_{R}=a$ or $p(R)=aR$. 
    Clearly, eliminating $\chi$ in exponential models where $p_{R}=ae^{\lambda R}$ is extremely difficult.
\end{itemize}
It appears that among the polynomial expansions of $p(R)$, Starobinsky-like model is a special case for which one can safely remove the non-dynamical DOF $\chi$ in the Einstein frame without imposing additional constraints on the scalar potentials. We shall, therefore, proceed with this choice for the rest of the analysis. Rewriting the action accordingly,

\begin{equation}\label{eq:palatini-action}
  S=\int d^{4}x\sqrt{-{g}} \left[\frac{{R}}{2\kappa} - \frac{1}{2(q+8\kappa \alpha V)} \partial_{\mu}{\phi}\partial^{\mu}{\phi} + \frac{\kappa \alpha }{2(q+8\kappa \alpha V)} (\partial{_\mu}\phi\partial{^\mu}\phi) (\partial{_\nu}\phi\partial{^\nu}\phi) -U \right],
\end{equation}
where,
\begin{equation}
    U\equiv\frac{V}{q (q+8\kappa \alpha V)},
\end{equation}
is the effective potential. Compared to metric formalism, where a wider variety of $f(R,\phi)$ models could give favorable conditions, it appears that Palatini formalism demands a smaller subset of analytic functions $p(R)$ in order for the non-dynamical scalar to be eliminated. Now that we have isolated these conditions solely from consistency requirements, we shall now verify whether \eqref{eq:palatini-action} in particular can support a proper cosmological evolution from early to late-times.

\section{Dynamical Stability Analysis of Einstein Frame Palatini $f(R,\phi)$ model}\label{sec:dynamical}
So far, we have left the potential $V$ and the factorized coupling function $q$ as arbitrary. We shall continue to do so in this section as well and check if we can find some conditions for both $V$ and $q$ that ensure proper late-time acceleration. We will, however, make one well-posed simplification in \eqref{eq:palatini-action} to make the forthcoming analysis easier.
Starobinsky inflation in metric formalism can be obtained from a general $f(R,\phi)$ theory by assuming that $q(\phi)=1$ and $p(R)=R+\alpha_mR^2$, where $\alpha_m\approx10^9\kappa$ from Planck 2018 \cite{Planck:2018} constraints on the total energy density of the universe during the inflationary epoch. In terms of $\alpha_m$, the total energy density during a slow-roll inflation in metric Starobinsky model can be expressed as $\approx (8\kappa\alpha_m)^{-1}$ \cite{Ellis:2015pla}. Since observational data is independent of the choice of model or formalism, if we assume that \eqref{eq:palatini-action} drives early-universe inflation instead, then during inflation,
\begin{equation}\label{eq:inflation-potential-planck}
    \frac{V}{q (q+8\kappa \alpha V)}\approx \frac{1}{8\kappa\alpha_m},
\end{equation}
where we have assumed that the slow-roll inflationary phase is driven by potential energy density domination. Now, considering the case where $q\to1$ in Palatini Einstein frame action, we can make the identification $\alpha\to\alpha_m$ if and only if,
\begin{equation}
    8\kappa \alpha V\gg q, 
\end{equation}
i.e. the scalar potential $V(\phi)$ ($\neq0$) doesn't contribute and inflation is driven solely by the $R^2$ term. Thus, Starobinsky inflation in Palatini formalism is expected to show similar behavior to metric formalism even in the absence of a dynamical scalaron mode \cite{Starobinsky:1980te}. Now, if we move away from a pure Starobinsky inflation such that $\phi$ provides a non-trivial contribution to inflationary dynamics and $q(\phi)>1$, the dimensionless quantity $\kappa\alpha V$ must undergo careful calibration to ensure that the condition \eqref{eq:inflation-potential-planck} remains true.

Post-inflation, however, $U$ must decrease significantly to allow further cosmological evolution. Keeping $\alpha$ constant during post-inflationary phases, we can claim without loss of generality that,
\begin{equation}
    \frac{8\kappa\alpha V}{q}\ll1,
\end{equation}
which is especially true during the present epoch where $U\leq 3H_0^2/\kappa$, and $H_0$ is the observed Hubble parameter at present time (the inequality suggests that the total energy density in the present epoch may not be dominated by the potential $U$ alone). We can, then, rewrite the action \eqref{eq:palatini-action} as,
\begin{equation}\label{eq:palatini-canonicalized-action-simplified}
     S=\int d^{4}x\sqrt{-{g}} \left[\frac{{R}}{2\kappa} - \frac{1}{2q} \partial_{\mu}{\phi}\partial^{\mu}{\phi} + \frac{\kappa \alpha }{2q} (\partial{_\mu}\phi\partial{^\mu}\phi) (\partial{_\nu}\phi\partial{^\nu}\phi) -U \right]+S_m, 
\end{equation}
where $U= V/q^2$ and $S_m$ represents non-relativistic dark matter. Note that the only remnant of the Jordan frame $R^2$ term is now the quartic-order kinetic coupling term. The Friedmann equations for this model containing both scalar field and a perfect fluid in FLRW background \eqref{eq:flrw-metric} are:
\begin{align}
    H^2&=\frac{\kappa}{3}\left(\frac{\dot{\phi}^2}{2q}+\frac{3\alpha \kappa \dot{\phi}^4}{2q}+U+\rho_{m} \right),\label{eq:friedmann-hsquare}\\
    \dot{H}&=-\frac{\kappa}{2}   \left(\frac{\dot{\phi}^2}{q}+\frac{2 \alpha  \kappa  \dot{\phi}^4}{q}+p_m+\rho _m\right). \label{eq:friedmann-hdot}
\end{align}
Here, $\rho_{m}$ represents the energy density
for dark matter (DM) taken as pressureless ($p_{m} = 0$) dust with
equation of state parameter $\omega_{m} = 0$. Now, by taking the variation of the action given in  \ref{eq:palatini-canonicalized-action-simplified} with respect to the scalar field $\phi$, we obtain the Klein-Gordon equation as,
\begin{equation}
\ddot{\phi}+3 H \dot{\phi}  \frac{(1+2 \alpha  \kappa  \dot{\phi}^2)}{(1+6 \alpha  \kappa  \dot{\phi}^2)}+\frac{V_{\phi}}{q(1+6 \alpha  \kappa  \dot{\phi}^2)}-\frac{q_{\phi} \dot{\phi}^2}{2q} \frac{(1+3 \alpha  \kappa  \dot{\phi}^2)}{(1+6 \alpha  \kappa  \dot{\phi}^2)}-\frac{2 q_{\phi} }{q^2} \frac{V}{(1+6 \alpha  \kappa  \dot{\phi}^2)}=0.\label{eq:phiddot}
\end{equation}
We define the following dimensionless variables,
\begin{equation}\label{eq:dynvar}
    x\equiv\frac{\kappa\dot{\phi}^2}{6 q H^2},\qquad y\equiv\frac{\alpha \kappa^2 \dot{\phi}^4}{2 q H^2},\qquad z\equiv\frac{\kappa U}{3H^2}, \qquad \lambda\equiv\frac{q_{\phi}}{q} \frac{\dot{\phi}}{H}.
\end{equation}
Using these, from  \eqref{eq:friedmann-hsquare}, we obtain the constraint equation:
\begin{equation}
    x+y+z+\Omega_{m}=1,
\end{equation}
where the relative energy density of matter $\Omega_{m}=\frac{\kappa \rho_{m} }{3 H^2 }$  is taken to be positive, such that,
\begin{equation}
    \Omega _{m}=1-x-y-z \leq 1.
\end{equation}
We also find expressions for energy density $\rho_{\phi}$ and pressure $p_{\phi}$ respectively as:
\begin{align}
    \rho_{\phi}&=\frac{\dot{\phi}^2}{2q}+\frac{3 \alpha \kappa \dot{\phi}^4}{2q}+U,\\
    p_{\phi}&=\frac{\dot{\phi}^2}{2q}+\frac{ \alpha \kappa \dot{\phi}^4}{2q}-U.
\end{align}
Now, we can obtain the relevant cosmological parameters using the dynamical variables we defined in  \eqref{eq:dynvar}. For the scalar field (DE), the density parameter and the equation of state can be respectively expressed as, 
\begin{align}
    \Omega _{\phi }&=\frac{\kappa  \rho _{\phi }}{3 H(t)^2}=(x+y+z), \\
    \omega _{\phi }&=\frac{p_{\phi }}{\rho _{\phi }}=\frac{3 x+y-3 z}{3 (x+y+z)}.
\end{align}
In addition, we can also express the effective equation of state parameter $\omega_{\rm eff}$ as,
\begin{equation}
\omega_{\rm eff}=\frac{p_{\phi}+p_{m}}{\rho_{\phi}+\rho_{m}}=x+\frac{y}{3}-z,
\end{equation}
Now, by using the Friedmann equations given in eqs. \eqref{eq:friedmann-hsquare}, \eqref{eq:friedmann-hdot} and the Klein-Gordon equation \eqref{eq:phiddot}, we can obtain the following dynamical system for our analysis,
\begin{align}
    x'&=2 I x+J x-\lambda  x, \label{eq:d1}\\
    y'&=4 I y+J y-\lambda  y, \label{eq:d2}\\
    z'&=J z-2 \lambda  z+z \sigma, \label{eq:d3}\\
    \lambda'&=I \lambda +\frac{J \lambda }{2}+\lambda ^2 \rho -\lambda ^2, \label{eq:d4}
\end{align}
where,
\begin{align}
    I &=\frac{-3 x-2 y}{x+2 y}+ \frac{\lambda  x+\lambda  y+2 \lambda  z-z \sigma }{2 (x+2 y)},\label{eq:d5}\\
    J &= 3 x+y-3 z+3,\label{eq:d6}\\
    \sigma& \equiv\frac{V_{\phi}}{V} \frac{\dot{\phi}}{H},\label{eq:d7}\\
    \rho &\equiv \frac{q\, q_{\phi\phi}}{q_\phi^2}.\label{eq:d8}
\end{align}
Here $\sigma$, $\rho$ are parameters, and the $'$ over the quantity signifies a corresponding derivative with respect to the number of e-folds $N$,  which can be related to cosmic time using the relation $dN=Hdt$. We will now solve the autonomous system (\ref{eq:d1}-\ref{eq:d4}) to find the fixed points and their corresponding eigenvalues, and also identify the conditions under which the model yields late-time accelerated attractors. The critical points along with their respective eigenvalues are given in Tables \ref{tab:xyzlambda_points} and \ref{tab:xyzlambda_eigen}.

\begin{table}[H]
\centering
\small
\begin{tabular}{|c|c|c|c|c|c|c|}
\hline
\textbf{Points} & \boldmath $x$ & \boldmath $y$ & \boldmath $z$ & \boldmath $\lambda$ & \boldmath $\Omega_{\phi}$ & \boldmath $\displaystyle \omega_{\rm eff}$ \\[4pt]
\hline
$P_1$ & $-2$ & $3$ & $0$ & $0$ & $1$ & $-1$ \\[6pt]
\hline
$P_2$ & $0$ & $1$ & $0$ & $\dfrac{4}{3-4\rho}$ & $1$ & $\dfrac{1}{3}$ \\[10pt]
\hline
$P_3$ & $0$ & $-\dfrac{\sigma}{4}$ & $\dfrac{\sigma+4}{4}$ & $0$ & $1$ & $-\dfrac{\sigma+3}{3}$ \\[12pt]
\hline
$P_4$ & $0$ & $\dfrac{(3-4\rho)\sigma}{16\rho-4}$ & $\dfrac{4\rho(\sigma+4)-3\sigma-4}{16\rho-4}$ & $\dfrac{\sigma}{4\rho-1}$ & $1$ & $\dfrac{3(\sigma+1)-4\rho(\sigma+3)}{3(4\rho-1)}$ \\[14pt]
\hline
$P_5$ & $0$ & $1$ & $0$ & $0$ & $1$ & $\dfrac{1}{3}$ \\[8pt]
\hline
$P_6$ & $1$ & $0$ & $0$ & $0$ & $1$ & $1$ \\[6pt]
\hline
$P_7$ & $-\dfrac{\sigma}{6}$ & $0$ & $\dfrac{\sigma+6}{6}$ & $0$ & $1$ & $-\dfrac{\sigma+3}{3}$ \\[12pt]
\hline
\end{tabular}
\caption{Critical points of the dynamical system with corresponding $\Omega_{\phi}$ and $\omega_{\rm eff}$.}
\label{tab:xyzlambda_points}
\end{table}

\begin{table}[H]
\centering
\small
\begin{tabular}{|c|c|c|}
\hline
\textbf{Points} & \textbf{Eigenvalues} & \textbf{Stability Condition} \\ \hline

$P_1$ &
$\{-3,\,-3,\,0,\,\sigma\}$ &
Stable if $\sigma<0$  \\ \hline

$P_2$ &
$\left\{\dfrac{4-8\rho}{3-4\rho},\,-1,\,1,\,\dfrac{4\rho(\sigma+4)-3\sigma-4}{4\rho-3}\right\}$ &
Unstable \\ \hline

$P_3$ &
$\left\{-\tfrac{\sigma}{2},\,-\tfrac{\sigma}{4},\,-\sigma-3,\,-\sigma-4\right\}$ &
Stable if $\sigma>0$ \\ \hline

$P_4$ &
\begin{minipage}{0.58\linewidth}
\[
\left\{
\frac{\sigma-2\rho\sigma}{4\rho-1},\;
\frac{3(\sigma+1)-4\rho(\sigma+3)}{4\rho-1},
\right.
\]
\[
\left.
\frac{-16\rho^2(3\sigma+16)+32\rho(2\sigma+7)-21\sigma-40\pm\sqrt{\Delta(\rho,\sigma)}}{8(4\rho-3)(4\rho-1)}
\right\}
\]
\end{minipage}
&
See Fig. \ref{fig:D4 plot}

 \\ \hline

$P_5$ &
$\{2,\,1,\,1,\,\sigma+4\}$ &
Unstable \\ \hline

$P_6$ &
$\{-6,\,3,\,0,\,\sigma+6\}$ &
Unstable  \\ \hline

$P_7$ &
$\{0,\,\sigma,\,-\sigma-3,\,-\sigma-6\}$ &
Stable if $-3<\sigma<0$ \\ \hline

\end{tabular}
\caption{Fixed points, eigenvalues, and stability conditions.}
\label{tab:xyzlambda_eigen}
\end{table}

\begin{figure}[ht!]
    \centering
\includegraphics[width=0.45\linewidth,height=0.3\linewidth]{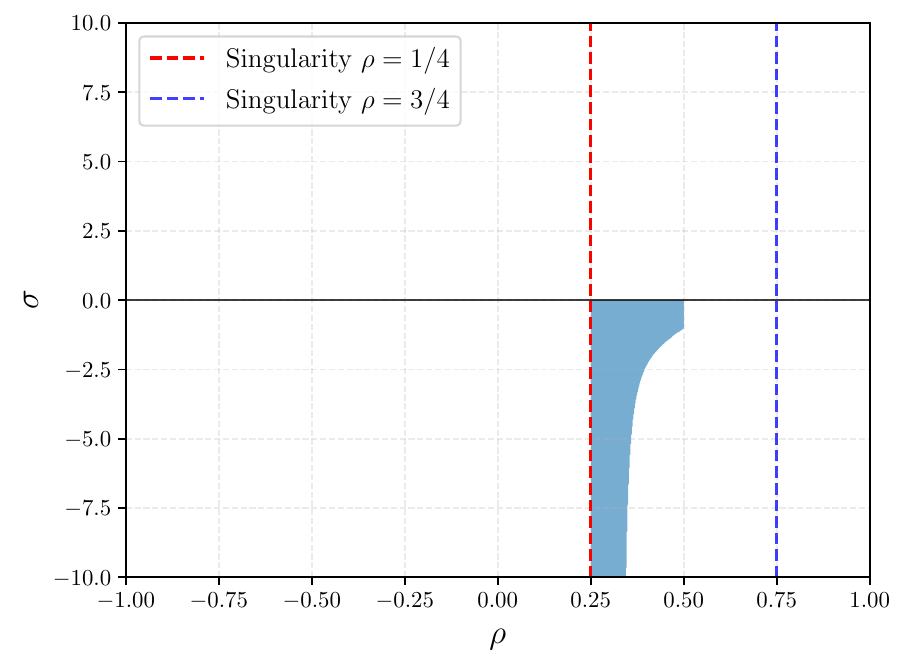}
    \caption{Point $P_4$ is stable in the given parameter space of $(\sigma,\rho)$. The dashed lines represent forbidden values of $\rho$ where the eigenvalues can be seen to diverge in Table \ref{tab:xyzlambda_eigen}.}
    \label{fig:D4 plot}
\end{figure}

\begin{figure}[ht!]
    \centering

    \begin{subfigure}{0.45\linewidth}
        \centering
        \includegraphics[width=0.9\linewidth,height=0.55\linewidth]{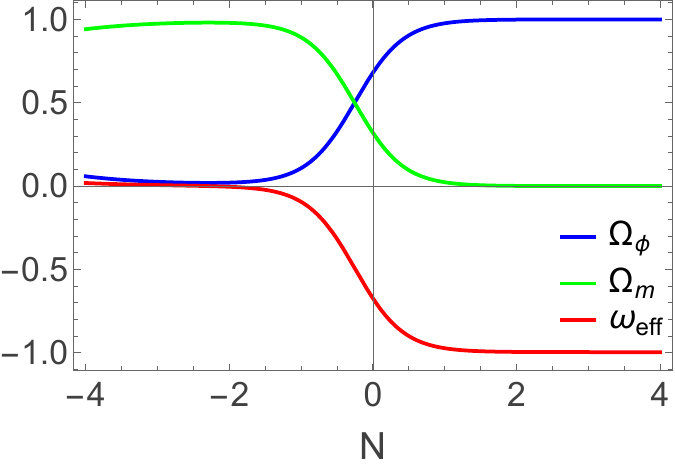}
        \caption{}
    \end{subfigure}
    \hspace{0.3cm}
    \begin{subfigure}{0.45\linewidth}
        \centering
        \includegraphics[width=0.9\linewidth,height=0.55\linewidth]{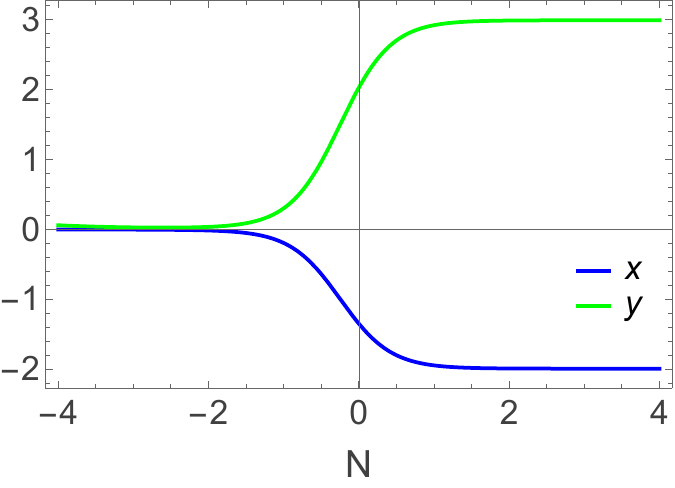}
        \caption{}
    \end{subfigure}

    \vspace{0.35cm}

    \begin{subfigure}{0.45\linewidth}
        \centering
        \includegraphics[width=0.9\linewidth,height=0.55\linewidth]{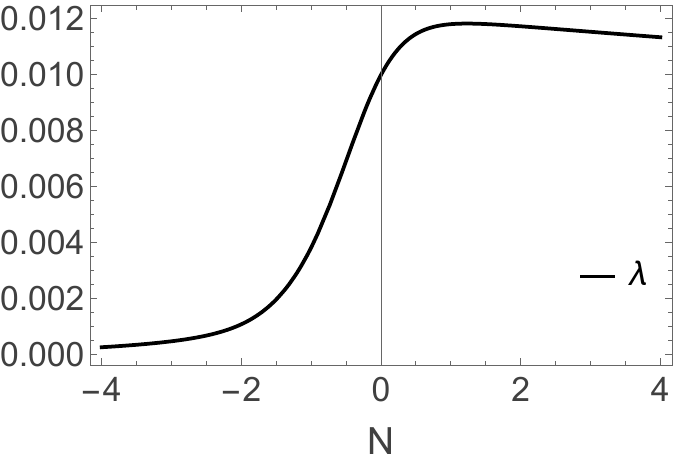}
        \caption{}
    \end{subfigure}
    \hspace{0.3cm}
    \begin{subfigure}{0.45\linewidth}
        \centering
        \includegraphics[width=0.9\linewidth,height=0.55\linewidth]{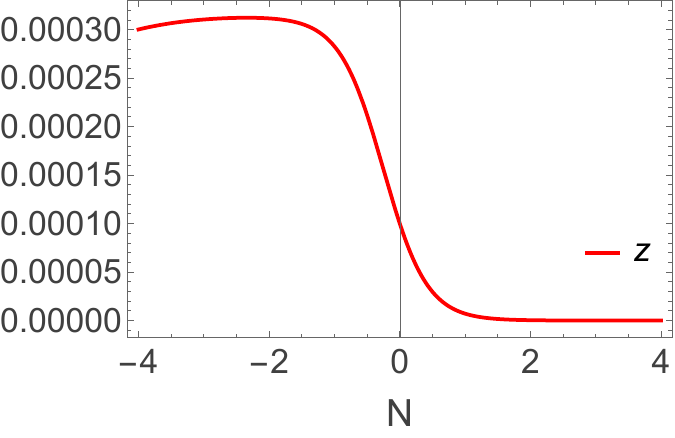}
        \caption{}
    \end{subfigure}

    \caption{Evolution of $\Omega_\phi$, $\Omega_m$, $\omega_{\rm eff}$, and the dynamical variables $x$, $y$,  $z$, and $\lambda$ for the initial conditions $\Omega_{\phi}(0)=0.68$, $\omega_{\phi}(0)=-0.99$, $z(0)=10^{-4}$, and $\lambda(0)=0.01$, with model parameters $\sigma=-1$ and $\rho=-1$. As the e-folding number $N$ increases, the scalar-field energy density $\Omega_\phi$ steadily grows and asymptotically approaches unity, while the matter component $\Omega_m$ diminishes to zero. Consequently, the effective equation of state $\omega_{\rm eff}$ evolves toward the cosmological-constant value $\omega_{\rm eff}\simeq -1$, signaling late-time accelerated expansion. The plots of the dynamical variables show a consistent evolution: $x$ decreases and stabilizes near $-2$, $y$ increases and settles around $3$, the potential-related variable $z$ approaches zero, and $\lambda$ remains extremely small throughout the evolution. For $\sigma=-1$, Tables~\ref{tab:xyzlambda_points} and \ref{tab:xyzlambda_eigen} indicate that both critical points $P_{1}$ and $P_{7}$ can correspond to late-time acceleration. However, the trajectories of the dynamical variables clearly converge to the critical point $P_{1}$, confirming it as the late-time attractor of the model.}
    \label{fig:xyzlambda_grid}
\end{figure}

\begin{figure}[H]
    \centering

    \begin{subfigure}{0.45\linewidth}
        \centering
        \includegraphics[width=\linewidth]{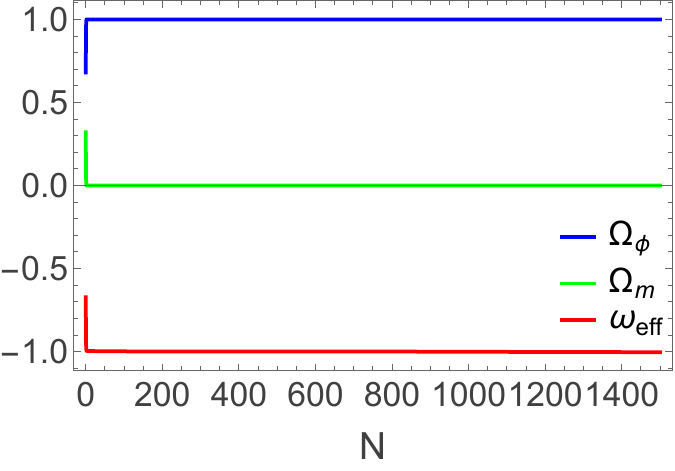}
        \caption{}
    \end{subfigure}
    \hspace{0.3cm}
    \begin{subfigure}{0.45\linewidth}
        \centering
        \includegraphics[width=\linewidth]{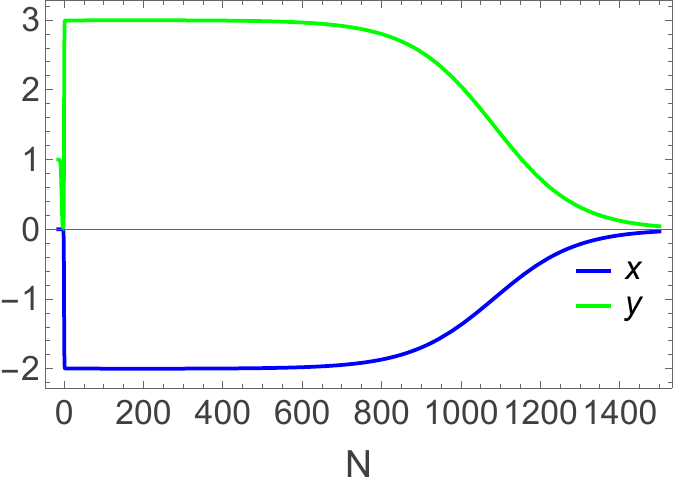}
        \caption{}
    \end{subfigure}

    \vspace{0.35cm}

    \begin{subfigure}{0.45\linewidth}
        \centering
        \includegraphics[width=\linewidth]{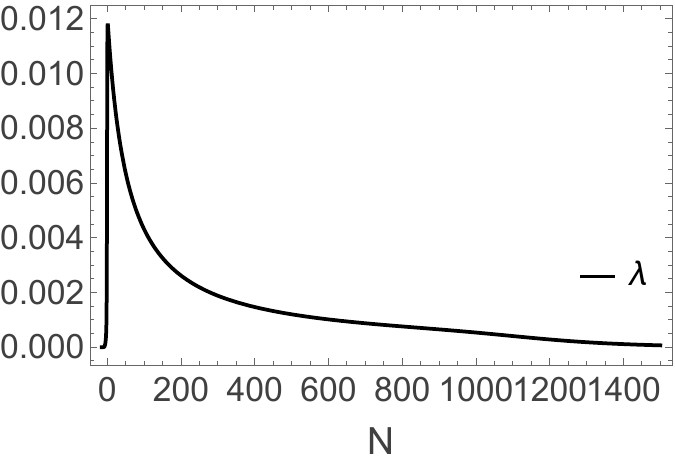}
        \caption{}
    \end{subfigure}
    \hspace{0.3cm}
    \begin{subfigure}{0.45\linewidth}
        \centering
        \includegraphics[width=\linewidth]{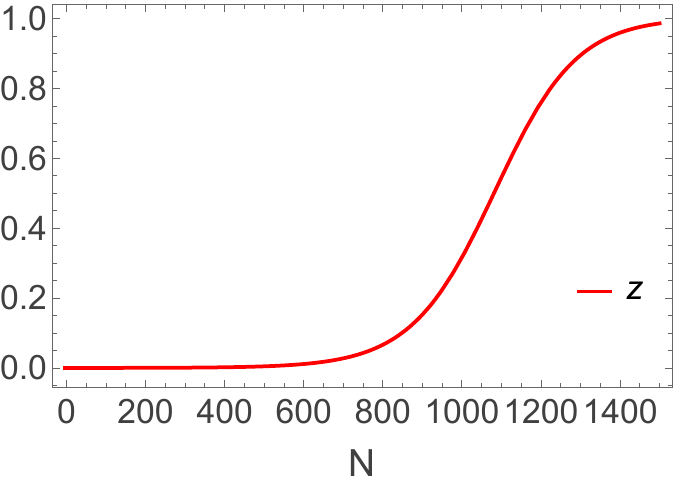}
        \caption{}
    \end{subfigure}

    \caption{Evolution of $\Omega_\phi$, $\Omega_m$, $\omega_{\rm eff}$, 
    $x$, $y$, $z$, and $\lambda$ with the number of e-folds $N$,
    for the initial conditions  
    $\Omega_{\phi}(0)=0.68$, $\omega_{\phi}(0)=-0.99$, 
    $z(0)=10^{-4}$, $\lambda(0)=0.01$, and the parameters 
    $\sigma = 0.01$, and $\rho = -0.1$. From panel (a), the scalar-field energy density parameter $\Omega_\phi$ approaches unity, while the matter density parameter $\Omega_m$ decays to zero, and the effective equation of state parameter converges to $\omega_{\rm eff}=-1.00333$, signaling a late-time phantom-like accelerated expansion. Panel (b) indicates that the dynamical variables $x$ and $y$ asymptotically decay towards vanishing values at late times. Panel (c) shows that the variable $\lambda$ decreases and tends to zero, whereas panel (d) illustrates that the variable $z$ grows and approaches to $z\simeq 1$ as $N$ increases. By comparing this asymptotic behavior with the critical points summarized in Table \ref{tab:xyzlambda_points} and Table \ref{tab:xyzlambda_eigen}, the system is seen to approach the fixed point $P_3$. For $\sigma>0$, this point is stable and therefore represents the late time dark energy dominated attractor of the system.}
    \label{fig:xyzlambda_case}
\end{figure}

\begin{itemize}
    \item \textbf{Critical Point $P_1$:}
    This point represents a de Sitter vacuum solution dominated by the scalar field ($\Omega_{\phi}=1$) with an effective equation of state $\omega_{\rm eff}=-1$. Its stability is determined by the parameter $\sigma$; specifically, it possesses a zero eigenvalue, making it non-hyperbolic. However, for $\sigma<0$, the centre manifold dynamics allow it to function as a stable late-time attractor, physically corresponding to a cosmological constant (this is also demonstrated in the dynamics of the reduced system; see Appendix). Alternatively, if $\sigma>0$, the point becomes a saddle implying that the current acceleration era may be transient and may lead to a different epoch.
    
    \item \textbf{Critical Point $P_2$:}
    The point $P_2$ is a saddle point as it has mixed eigen values and with equation of state parameter $\omega_{\rm eff}=\frac{1}{3}$, this point cannot serve as a late-time attractor and represents only a transient epoch in the cosmic history.
    \item \textbf{Critical Point $P_3$:}
    This point represents a phantom dark energy attractor ($\omega_{\rm eff}<-1$) whenever $\sigma>0$. In this parameter region, all eigenvalues are negative, implying that $P_3$ a fully stable and offers a mathematically robust model for phantom acceleration as the final state of the universe. However, for $\sigma \le 0$, it becomes unstable, thus losing its viability.
    \item \textbf{Critical Point $P_4$:}
    This is the most dynamically complex point in the system, with an effective equation of state and stability dependent on both $\rho$ and $\sigma$. It can behave as a stable attractor, a repeller, or a saddle depending on parameters $\rho$ and $\sigma$. We illustrate the stability region of the fixed point $P_4$ in the $(\sigma,\rho)$ parameter space in Fig. \ref{fig:D4 plot}, where,
    \begin{align}
        \Delta(\rho,\sigma)=&\ 256 \rho^{4} (5\sigma+16)^{2}-1024 \rho^{3}\!\left(16\sigma^{2}+85\sigma+112\right)+32\rho^{2}\!\left(479\sigma^{2}+2056\sigma+2208\right)\nonumber\\
        &-64\rho\!\left(96\sigma^{2}+323\sigma+280\right) +873\sigma^{2}+2256\sigma+1600.
    \end{align}
    \item\textbf{Critical Point $P_5$:}
    $P_5$ can act as saddle or unstable depending on parameter $\sigma$. Physically, $P_5$ can behave as  an early-time repeller if $\sigma > -4$ (all positive eigenvalues), and ensures the universe evolves away from this point very quickly. For other choices of $\sigma$, the point will act as saddle that would represent a transient phase.    
    \item \textbf{Critical Point $P_6$:}
    This point corresponds to stiff fluid phase where $\omega_{\rm eff}=1$. The eigenvalue spectrum includes a positive value, indicating that $P_6$ is unstable. While such stiff-matter solutions often appear as transient phases in the early universe, this point cannot support late-time acceleration. It functions dynamically as a saddle, bridging earlier cosmological epochs but never serving as a final attractor.
    \item \textbf{Critical Point $P_7$:}
    It offers a phenomenologically promising alternative to the phantom point $P_3$. It possesses the same effective equation of state form but distinct stability properties. For $-3<\sigma<0$, it acts as a stable, accelerating attractor with quintessence-like behavior ($-1<\omega_{\rm eff}<-1/3$). This makes $P_7$ another strong candidate for describing observed cosmic acceleration.
\end{itemize}
Figures \ref{fig:xyzlambda_grid} and \ref{fig:xyzlambda_case} can be shown to represent two distinct scenarios corresponding to points $P_1$ and $P_3$ respectively depending on the parameters $\sigma$ and $\rho$. Please note that even though some points appear non-hyperbolic (with one or more zero eigenvalues), their stability has been verified in the reduced parameter space (see Appendix). Also note that even though $\rho$ and $\sigma$ may appear to be dynamical, their treatment here is as constant parameters. This is because if they are considered dynamical, then $\rho'$ and $\sigma'$ would contain higher derivatives of potentials $V(\phi)$ and $q(\phi)$ with respect to $\phi$ which would require us to define variables for successive derivatives of these potentials to study their evolution. However, $\rho$ and $\sigma$ being positive or negative near a fixed point could have profound implications on the behavior of $V(\phi)$ and $q(\phi)$.

The analysis performed so far clearly establishes the theoretical consistency of the model and demonstrates the existence of a stable late-time accelerating solution. However, this analysis alone is qualitative and does not determine whether the predicted expansion history agrees with observations. To test the observational viability of the model and constrain its parameters, we now confront it with late-time cosmological data in the next section.


\section{Data Analysis}\label{sec:data-analysis}
Our subsequent objective requires performing data analysis to constrain the parameter space we have in the model. For achieving this, deriving an analytical expression for the Hubble parameter as a function of redshift is necessary. While this process is quite straightforward for the $\Lambda$CDM model, it poses challenges in our $k$-essence model owing to the inability to obtain an analytical solution to the Friedmann equation. Therefore, we opt for a dynamical systems approach employing the dynamical variables obtained in section \ref{sec:dynamical}. Solving these coupled differential equations, we capture the evolution of the Hubble parameter. In our analysis, we use the following publicly available late time datasets,
\begin{enumerate}
     \item \textbf{DESI:-} We use the 13 DESI-BAO DR2 (\cite{DESI:2025zgx,DESI:2025zpo}) measurements across the redshift range $0.1 < \textbf{z} < 4.2$ obtained from observations of about 14 million galaxies and quasars which include bright galaxy sample (BGS), luminous red galaxies (LRG), emission line galaxies (ELG), quasars (QSO), and Lyman-$\alpha$ tracers. These measurements are given in terms of the volume averaged distance $D_{\rm V}(\textbf{z}) / r_{\rm d}$, angular diameter distance $D_{\rm M}(\textbf{z}) / r_{\rm d}$ and comoving Hubble distance $D_{\rm H}(\textbf{z}) / r_{\rm d}$, where $r_{\rm d}$ is the sound horizon at the drag era. 
     \item \textbf{Cosmic Chronometers:-} The Cosmic chronometers (CC) approach allows us to obtain observational values of the Hubble function at different redshifts $\textbf{z}\leq2$ directly. Since these measurements are independent
     of any cosmological model and Cepheid distance scale, they can be used to place better constraints on it. In the present analysis, we measure $H(\textbf{z})$ using the CC covariance matrix \cite{2012JCAP...08..006M,2015MNRAS.450L..16M,Moresco:2016mzx}
     \item \textbf{SNeIa:-} We use the PantheonPlus (PP) dataset \cite{Scolnic:2021amr}, which contains 1701 light curves for 1550 spectroscopically confirmed Type Ia supernovae (SNeIa) covering the redshift range $0.001 < \textbf{z} < 2.26$. Additionally, we incorporate the Union3 compilation comprising 2087 SNe \cite{Rubin:2023ovl}. We also use the DESY5 sample comprising 1635 photometrically classified SNe from the released part of the full 5 year data of the Dark Energy Survey collaboration (with redshifts in the range 0.1 $< \ \textbf{z}\ <$ 1.3), complemented by 194 low-redshift SNe from the CfA3 \cite{2009ApJ...700..331H}, CfA4 \cite{2012ApJS..200...12H}, CSP \cite{Krisciunas:2017yoe}, and Foundation \cite{Foley:2017zdq} samples (with redshifts in the range 0.025 $< \ \textbf{z}\ <$ 0.1), for a total of 1829 SNe \cite{DES:2024jxu}. 
\end{enumerate}
To constrain the free and derived parameters in our model, we perform Markov-Chain-Monte Carlo (MCMC) simulations using the publicly available tool \texttt{COBAYA} \cite{Torrado:2020dgo,2019ascl.soft10019T}. We will be analysing the general structure of the dynamical equations obtained in eqs. (\ref{eq:d1}-\ref{eq:d4}). Further, we rewrite them, by replacing the dynamical variables $x$ and $y$ with cosmologically relevant quantities such as $\Omega_\phi$ and $\omega_\phi$. Employing the numerical integration routines in \texttt{Scipy} \cite{2020SciPy-NMeth}, we solve the dynamical system starting from $\ln a  = -4$, and evolve it to the present time. Thus our parameter space consists of the initial values of the four dynamical variables $\Omega_\phi^i$, $\omega_\phi^i$, $z_i$ and $\lambda_i$, and the parameters $\rho$ and $\sigma$ that appear in their dynamical equations. Further, we are taking $r_d$, the comoving sound horizon at the drag epoch, as a free parameter for analysing the BAO data from DESI. We constrain the parameters with uniform priors: $\Omega_\phi^i \in [0, 0.3]$, $\omega_\phi^i \in [-1,1]$, $z_i \in [10^{-6}, 10^{-3}]$, $\lambda_i \in [10^{-8},1]$, $\rho \in [-10,2]$, $\sigma \in [-10,2]$ and $r_d \in [100, 200]$. The convergence of chains is ensured by having the Gelman-Rubin criterion $|R-1| \leq 0.01$. We utilize  \texttt{GetDist} \cite{Lewis:2019xzd} and \texttt{BOBYQA} \cite{2018arXiv180400154C,2018arXiv181211343C} to analyze and minimize the chains. The marginalised 1-D and 2-D posterior distributions of the dynamical variables and other parameters in our model for some of the data set combinations are shown in Fig. (\ref{fig:contour}). In Table \ref{tab:constraints}, we report the marginalized parameter constraints along with their $1-\boldsymbol{\sigma}$ errors.
\begin{table}[h]
    \centering
    \begin{tabular}{|l|c|c|c|c|c|c|c|}
    \hline
      \textbf{Dataset}   & $ \Omega_\phi^i$ & $\omega_\phi^i$ & $\rho$ & $\sigma$ & $r_d$ & $H_0$ & $\Omega_m^0$\\
     \hline
      CC+DESI   & $0.1077^{+0.0082}_{-0.058}$ & $-0.23^{+0.53}_{-0.26}$& $< -4.98$ & $< -3.01$ & $140.5\pm 1.9$ & $71.90\pm 0.89$& $0.2928\pm 0.0083$ \\
      CC+DESI+PP   & $0.114^{+0.013}_{-0.061}$ & $-0.25\pm 0.38$ & $-3.37^{+2.6}_{-0.78}$ & $-4.5\pm3.1$ & $139.3\pm 1.9$ & $71.60\pm 0.88$ & $0.2970\pm 0.0088$\\
      CC+DESI+DESY5   &  $0.118^{+0.015}_{-0.061}$&  $-0.25^{+0.53}_{-0.23}$ & $-2.71^{+2.1}_{-0.46}$ & $-4.4^{+4.6}_{-5.5}$ & $138.9\pm 1.9$ & $71.51\pm 0.88$ & $0.2976\pm 0.0092$ \\
      CC+DESI+Union3   & $0.1088^{+0.0099}_{-0.058}$ & $-0.24^{+0.53}_{-0.22}$& $-4.9^{+3.5}_{-1.9}$& $<-2.95$ & $139.9\pm 1.9$ & $71.76\pm 0.89$ & $0.2961\pm 0.0083$ \\
      CC+DESI+PP+DESY5   & $0.127^{+0.018}_{-0.067}$ &  $-0.24^{+0.52}_{-0.22}$ & $-1.82^{+1.2}_{-0.17}$ & $-4.2\pm3.2$ &  $138.5\pm 1.8$ & $71.42\pm 0.86$ & $0.2961\pm 0.0095$ \\
      CC+DESI+PP+Union3   & $0.122^{+0.016}_{-0.066}$ & $-0.24^{+0.53}_{-0.34}$ & $-2.45^{+1.8}_{-0.34}$ & $-4.4\pm 3.1$ & $138.9\pm 1.9$ & $71.52\pm 0.87$ & $0.2965\pm 0.0092$ \\
      CC+DESI+Union3+DESY5 \hspace{1em}  & $0.123^{+0.018}_{-0.064}$ & $-0.25^{+0.52}_{-0.31}$ & $-2.12^{+1.5}_{-0.28}$ & $< -2.24$ & $138.6\pm 1.8$ & $71.45\pm 0.86$ &  $0.2971\pm 0.0094$ \\
      CC+DESI+Union3+DESY5+PP  & $0.130^{+0.018}_{-0.068}$ & $-0.24^{+0.53}_{-0.22}$ & $-1.72^{+1.2}_{-0.090}$ & $-4.2^{+4.5}_{-5.1}$ & $138.4\pm 1.8$ & $71.36\pm 0.88$ &  $0.2955\pm 0.0095$ \\
      \hline
    \end{tabular}
    \caption{The mean $\pm 1-\boldsymbol{\sigma}$ constraints on  cosmological parameters inferred from various datasets including CC, DESI DR2, and supernovae and their combinations. Here, $H_0$ is in units of km ${\rm s}^{-1}$ ${\rm Mpc}^{-1}$. The dynamical variables $z_i$ and $\lambda_i$ do not get constrained by the dataset combinations and hence are not reported in the table.} 
    \label{tab:constraints}
\end{table}
\begin{figure}[ht!]
    \centering

    \begin{subfigure}[t]{0.48\textwidth}
        \centering
        \includegraphics[width=0.9\linewidth,height=0.55\linewidth]{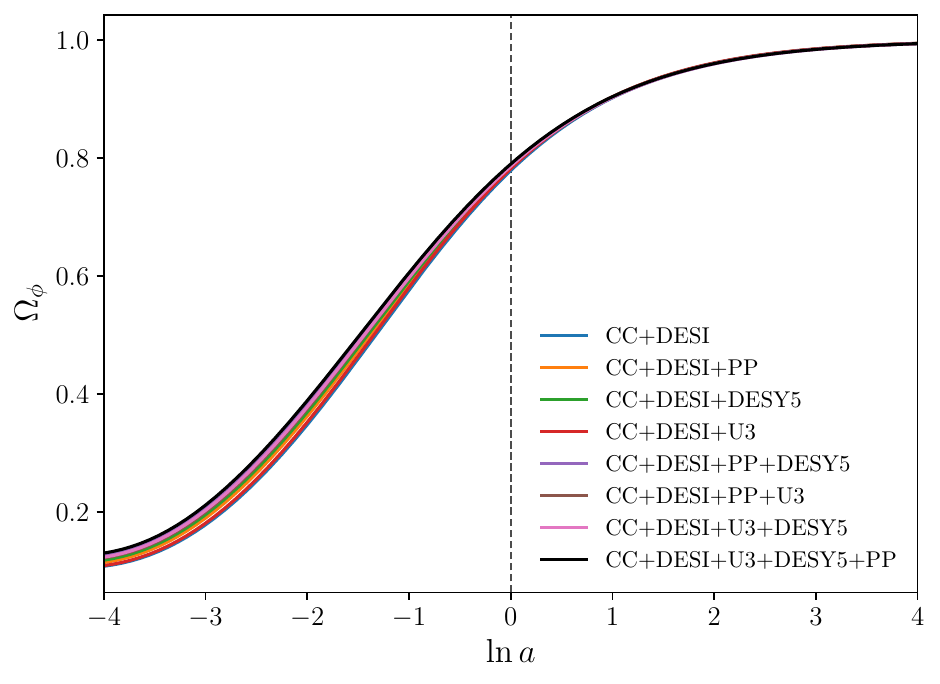}
       \caption{}
        \label{fig:omega_phi}
    \end{subfigure}
    \hfill
    \begin{subfigure}[t]{0.48\textwidth}
        \centering
        \includegraphics[width=0.9\linewidth,height=0.55\linewidth]{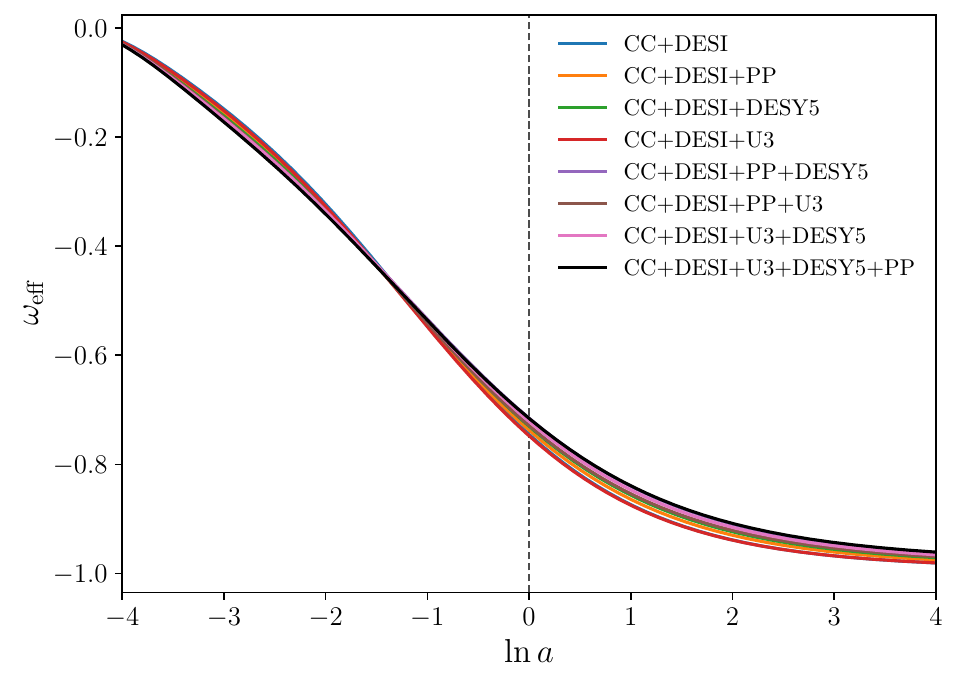}
       \caption{}
        \label{fig:w_eff}
    \end{subfigure}

    \vspace{0.5cm}

    \begin{subfigure}[t]{0.48\textwidth}
        \centering
        \includegraphics[width=0.9\linewidth,height=0.6\linewidth]{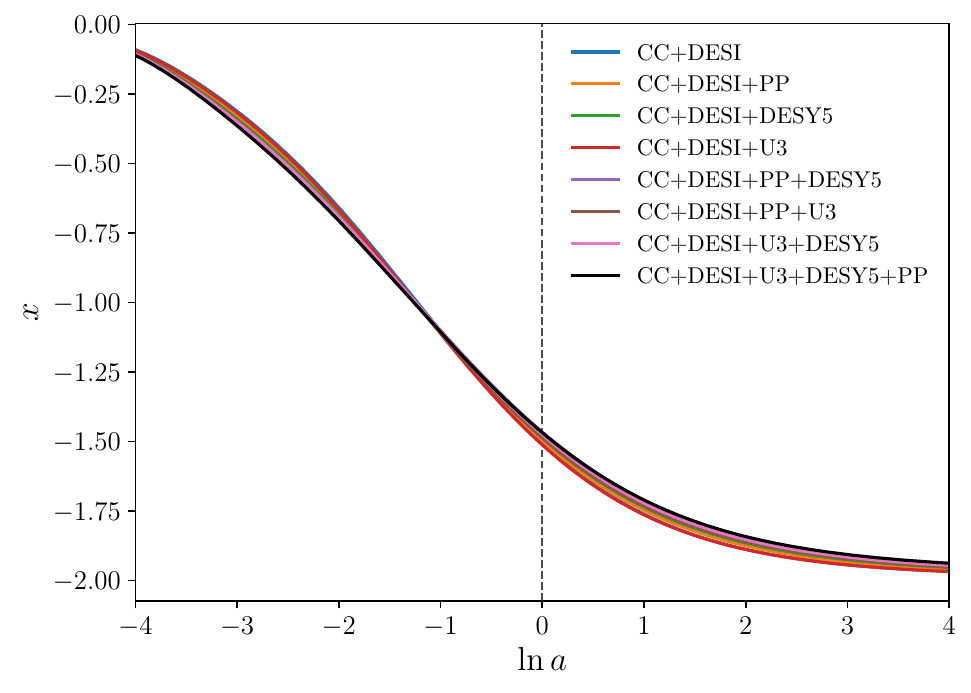}
       \caption{}
        \label{fig:X}
    \end{subfigure}
    \hfill
    \begin{subfigure}[t]{0.48\textwidth}
        \centering
        \includegraphics[width=0.9\linewidth,height=0.6\linewidth]{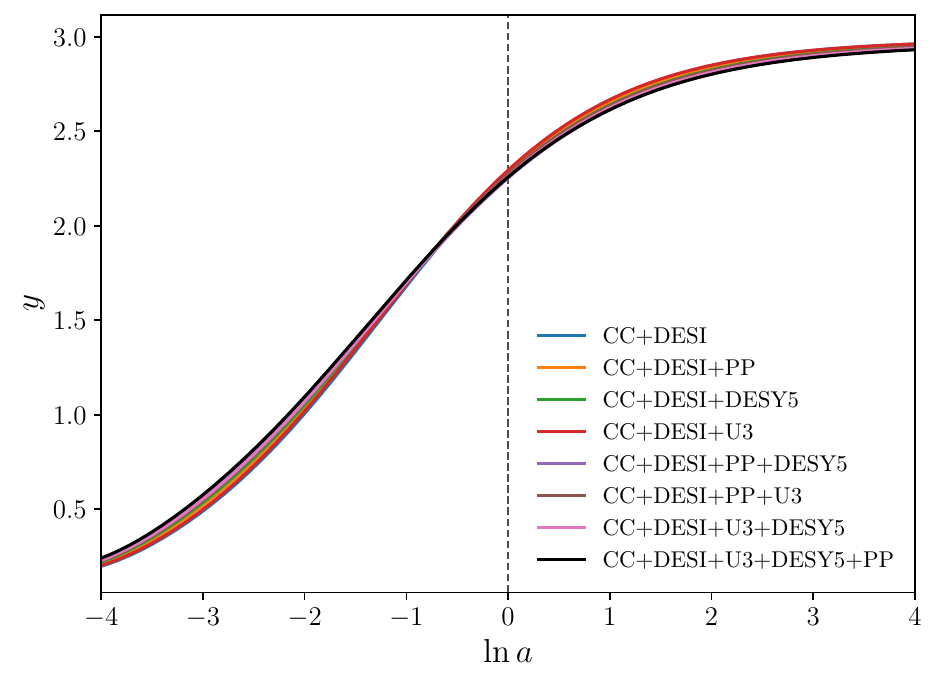}
        \caption{}
        \label{fig:Y}
    \end{subfigure}

    \caption{ Panels (a) and (b) respectively show the evolution of the field energy density and the effective equation of state.
    Panels (c) and (d) display the evolution of kinetic and hyper-kinetic terms of the field.
    The vertical dashed line denotes the present epoch.
    These plots are generated in accordance with the constraints listed in Table~\ref{tab:constraints}.}
    \label{fig:model_params_combined}
\end{figure}
We can see from Table \ref{tab:constraints} that the combination of DESI with CC can put only a bound on the model parameter $\rho$. However, with the inclusion of a supernova dataset to this combination, $\rho$ can be better constrained in our framework. The same can be identified from the corner plot given in Fig. (\ref{fig:contour}). Constraints can be obtained on the other model parameters as well. However, the parameters $z_i$ and $\lambda_i$ do not show any constraints and so we haven't reported them in our results. The evolution of the cosmological parameters in our $k$-essence model, such as the fractional energy density and the effective equation of state, in accordance with the constraints obtained from various dataset combinations, is shown in Fig. \ref{fig:model_params_combined}. In this plot, the evolution is depicted against the number of e-folds $N=\ln a$. The results demonstrate that the model exhibits a matter-dominated phase (i.e., $\Omega_m>\Omega_\phi$), for all the datasets, before transitioning to an accelerating epoch. During this matter dominated phase, the effective equation of state is close to zero, indicating attractive scaling behavior. As the system enters the late-time phase, the model shows accelerated expansion with $\omega_{\rm eff}<-1/3$. Further, $\omega_{\rm eff}$ is seen to approach to a value of -1 asymptotically in the future.
\begin{figure}[ht!]
    \centering
    \includegraphics[width=0.9\linewidth,height=0.9\linewidth]{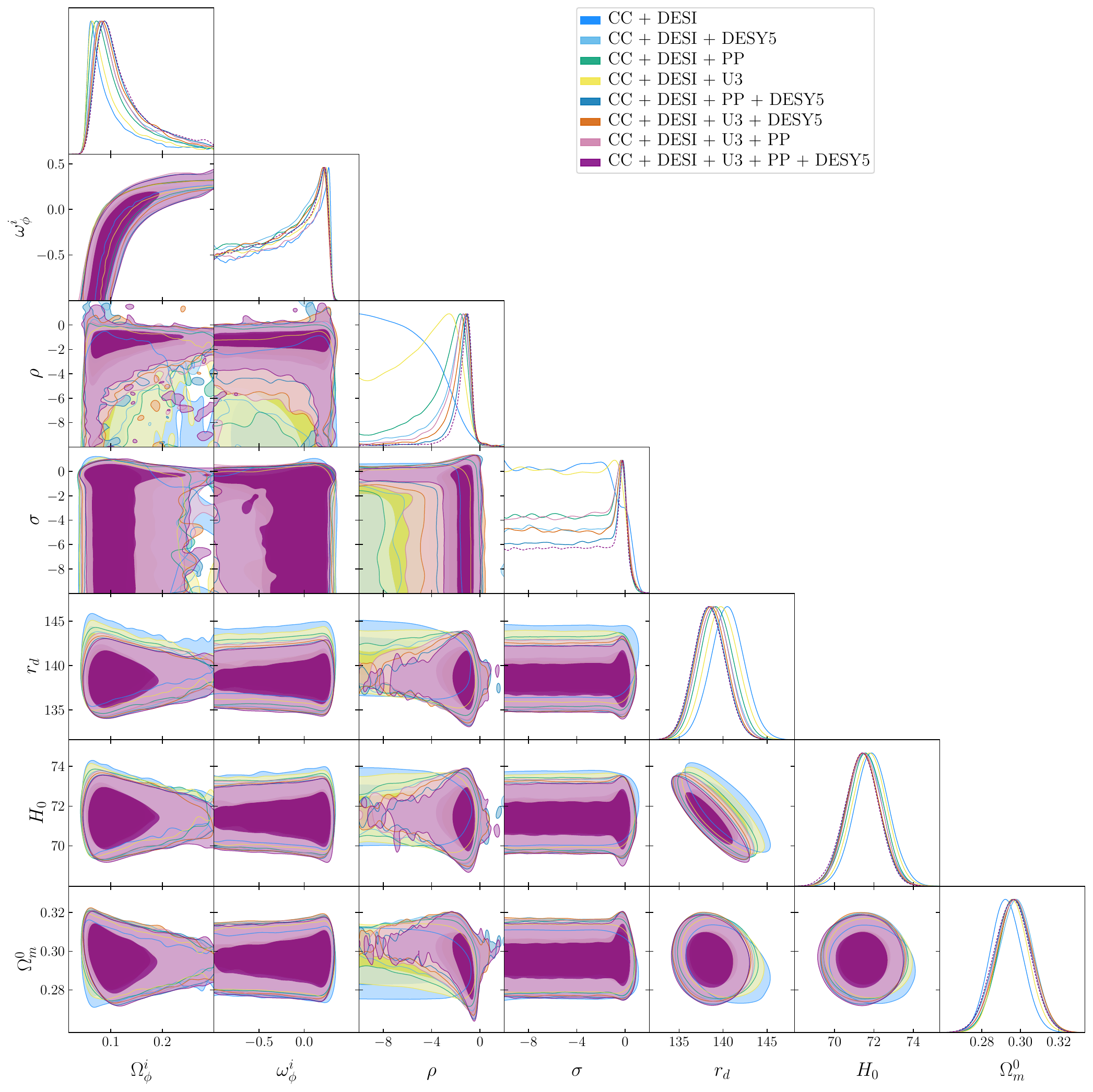}
    \caption{Corner plots of 1D and 2D marginalized posterior distributions of the model parameters in the presence of dataset combinations based on BAO from DESI DR2, Cosmic chronometers, and supernovae datasets. Contours at 68\% ($1-\boldsymbol{\sigma}$) and 95\% ($2-\boldsymbol{\sigma}$) levels
showing parameter constraints and correlations within the framework.}
    \label{fig:contour}
\end{figure}

\section{Discussion}\label{sec:discussion}
$f(R,\phi)$ gravity offers rich phenomenology and can even explain early- and late-time cosmic evolution under one unified umbrella. In most literature, $f(R,\phi)$ theories are studied by assuming specific form factors. In this paper, however, we attempted to constrain Palatini $f(R,\phi)$ theories in the Einstein frame based on consistency requirements, dynamical stability at late-time, as well as observational data to find compatibility with various parameters. We have used late time data sets to capture the behaviour of the cosmological and other parameters we have in our model. We have used a general potential in our analysis where it appear as a dynamical variable while its derivatives remain as parameters. With the observational data, we could not obtain robust constraints on all the parameters. However, we were able to constrain some of them, while obtaining bounds for the others. Particularly, the estimation of the parameters $\rho$ and $\sigma$ are important since they determine the stability of the fixed points in the dynamical system, thereby controlling the evolution. 

We found that among the fixed points listed in Table \ref{tab:xyzlambda_points}, only $P_1$ and $P_3$ provide a stable late-time acceleration era with behavior summarized in Table \ref{tab:xyzlambda_eigen} and plotted explicitly in Figs \ref{fig:xyzlambda_grid} and \ref{fig:xyzlambda_case} respectively. Now, as mentioned earlier, $\rho$ and $\sigma$ are treated as constants throughout the fixed point analysis and their values are later fixed at $N=-4$ using observational data in Table \ref{tab:constraints}. Within 68\% CL, we see that $\rho<0$. Based on its definition and knowing that $q\to1$ is the minimum allowed value to maintain consistency with general relativity, we find that $q_{\phi\phi}<0$. Also, even though observations seem to favor a negative $\sigma$ (i.e. $V_\phi<0$), small positive values may still be allowed.

If we consider $\sigma<0$, we rule out $P_3$ as a suitable fixed point since it demands $\sigma>0$. Below we list the various parameter values for $P_1$ as a stable fixed point and their corresponding implications:
\begin{align}
    x=-2,\ y=3& \qquad\implies\qquad \dot{\phi}^2\to-(24\alpha \kappa)^{-1},\label{eq:coupling-asymp-max}\\
     z=0& \qquad \implies\qquad U= Vq^{-2}\to0,\label{eq:potential-asymp-zero}\\
    \lambda=0&\qquad\implies\qquad q_\phi\to0,\label{eq:coupling-asymp-constant}\\
    \sigma<0&\qquad\implies\qquad V_\phi<0,\label{eq:potential-slope-asymp-negative}\\
    \rho<0&\qquad\implies\qquad q_{\phi\phi}<0.\label{eq:coupling-slope-slope-asymp-negative}
\end{align}
Moving step by step, conditions \eqref{eq:coupling-asymp-constant} and \eqref{eq:coupling-slope-slope-asymp-negative} imply at the coupling potential $q(\phi)$ approaches its maximum value at $P_1$. The scalar potential $V(\phi)$ is found to decrease in magnitude as $\phi$ increases, as shown in \eqref{eq:potential-slope-asymp-negative} and it approaches zero as per \eqref{eq:potential-asymp-zero}. Also, the combination $x<0$ and $y>0$, given their definitions in \eqref{eq:dynvar} implies that $\dot{\phi}^2<0$ which is characteristic of phantom fields (also implied by the effective equation of state parameter in Fig. \ref{fig:xyzlambda_grid}). One could argue that we could consider that $\dot \phi^2$ is positive and $\alpha<0$ instead to eliminate the phantom menace. This, however, would mean that for $x=-2$ and $y=3$, $q(\phi)$ would have to be negative. In Einstein frame, this would imply that the signs of the kinetic term and quartic kinetic coupling in the scalar part of action \eqref{eq:palatini-action} are flipped, leading to $\phi$ becoming a ghostly DOF\footnote{A negative $q(\phi)$ in the Jordan frame action would imply that gravity becomes repelling at late-times.}. Since, instabilities are, therefore, present in the system nonetheless, $\dot\phi^2<0$ is a safer choice.

If, on the other hand, we consider $\sigma\gtrsim0$ (which is allowed within 68\% CL), the viable late-time attractor would be $P_3$. Below we list its corresponding parameter values along with their implications:
\begin{align}
    x=0,\ y=0& \qquad\implies\qquad \dot{\phi}^2\to0\qquad\implies\qquad\dot\phi\to0^{\pm},\label{eq:phi-dot-square-zero}\\
     z=1& \qquad \implies\qquad U= Vq^{-2}\to 3M_{\rm Pl}^2H^2,\label{eq:potential-asymp-max}\\
    \lambda=0&\qquad\implies\qquad q_\phi\to0,\label{eq:coupling-asymp-constant2}\\
    \sigma\gtrsim0&\qquad\implies\qquad V_\phi>0\ \text{and}\ \dot\phi\gtrsim0,\quad \text{or}\quad V_\phi<0\ \text{and}\ \dot\phi\lesssim0,\label{eq:sigma-positive}\\
    \rho<0&\qquad\implies\qquad q_{\phi\phi}<0.\label{eq:coupling-slope-slope-asymp-negative2}
\end{align}
Here, conditions \eqref{eq:coupling-asymp-constant2} and \eqref{eq:coupling-slope-slope-asymp-negative2} again imply that $q$ is approaching its maximum value, as for $\sigma<0$. From condition \eqref{eq:sigma-positive}, we find two completely different cases: either $\phi$ slowly increases in magnitude with time and $V$ increases in magnitude with $\phi$ or vice versa. In both cases, we can at least guarantee that $V$ is directly proportional to $\phi$. Since condition \eqref{eq:potential-asymp-max} implies that the scalar potential $V(\phi)$ is approaching its maximum value $3q^2M_{\rm Pl}^2H^2$ at late times, the only viable combination is $V_\phi>0$ and $\dot\phi\gtrsim0$.

An important thing to note is that the scale of $N$ in Fig. \ref{fig:xyzlambda_case} is much larger than the scale of $N$ in Fig. \ref{fig:xyzlambda_grid}. From Table \ref{tab:xyzlambda_eigen}, one can see that $P_1$ could behave as a saddle point for $\sigma>0$. As such, Figs \ref{fig:xyzlambda_grid} and \ref{fig:xyzlambda_case} could imply that $P_1$ is simply a transient stage before the universe settles into $P_3$ at late-times. We reiterate that throughout the analysis we have treated $\sigma$ as a constant parameter due to restrictions in finite-dimensional dynamical systems analysis. If $\sigma$ is treated as a dynamical variable, it may eventually change behavior from $\sigma<0$ (as seen from data analysis performed in Section \ref{sec:data-analysis} for $N=-4$) to $\sigma>0$ much later than the present epoch. We cannot rule out this conclusion with absolute certainty since it requires us to define an infinite-dimensional dynamical system for general potentials $V$ and $q$. This is beyond the scope of the present work, but it could be pursued in the future as a follow-up.

Alternatively, if one assumes specific forms of the potentials $V(\phi)$ and $q(\phi)$, such that they conform to the behavior exhibited by points $P_1$ and $P_3$, the number of variables could become finite and the system could be solved using the methods presented in this work. But since we sought to study the behavior of the general $f(R,\phi)$ theories with arbitrary potentials in this paper, we shall leave that analysis as a future work as well.

\begin{acknowledgments}
    AV also acknowledges the Council of Scientific \& Industrial Research (CSIR), India for support under the Research Associateship program. SP is partially supported by the DST (Govt. of India) Grant No. SERB/PHY/2021057.
\end{acknowledgments}

\appendix
\section*{Appendix}
Here, we present dynamical evolution of the system described in Section \ref{sec:dynamical} under various constraints. This analysis is intended to demonstrate that the points that appear non-hyperbolic in Table \ref{tab:xyzlambda_eigen} can become hyperbolic in the constrained parameter space.
\subsection{$V(\phi)=0$ and $\lambda$ as Constant}
The autonomous system (\ref{eq:d1}-\ref{eq:d5}) admits three critical points in the $(x,y)$ phase space, which are given in Table-\ref{tab:xypoints} and corresponding eigenvalues and stability conditions are given in Table-\ref{tab:xyeigenval}. In the following, we analyse the stability and physical implications of each point separately.
\begin{table}[h!]
\centering
\begin{tabular}{|c|c|c|c|c|}
\hline
\textbf{Point} & \textbf{$x$} & \textbf{$y$} & \textbf{$\Omega_{\phi}$} & \textbf{$\omega_{\rm eff}$} \\
\hline
$A_{1}$ & $0$ & $1$ & $1$ & $\dfrac{1}{3}$ \\
\hline
$A_{2}$ & $\dfrac{\lambda - 4}{2}$ & $3 - \dfrac{\lambda}{2}$ & $1$ & $\dfrac{\lambda}{3} - 1$ \\
\hline
$A_{3}$ & $1$ & $0$ & $1$ & $1$ \\
\hline
\end{tabular}
\caption{Fixed points of the dynamical system along with the corresponding values of the scalar-field density parameter $\Omega_{\phi}$ and effective equation of state $\omega_{\rm eff}$.}
\label{tab:xypoints}
\end{table}
\begin{table}[h!]
\centering
\begin{tabular}{|c|c|c|}
\hline
\textbf{Point} & \textbf{Eigenvalues} & \textbf{Stability} \\
\hline
$A_{1}$ & $\left\{\;1,\; 2 - \dfrac{\lambda}{2}\;\right\}$ &
Unstable for $\lambda < 4$, \; saddle for $\lambda > 4$ \\
\hline
$A_{2}$ & $\left\{\;\lambda - 3,\; \dfrac{(\lambda - 4)(\lambda - 6)}{\lambda - 8}\;\right\}$ &
Stable for $\lambda < 3$ \;  \\
\hline
$A_{3}$ & $\left\{\;3,\; \lambda - 6\;\right\}$ &
Saddle for $\lambda < 6$, \; unstable for $\lambda > 6$ \\
\hline
\end{tabular}
\caption{Eigenvalue structure and dynamical stability of the fixed points of the system.}
\label{tab:xyeigenval}
\end{table}
\begin{itemize}
    \item \textbf{Critical Point $A_{1}$:} The fixed point $A_{1}$ corresponds to a scalar field dominated state with $\Omega_{\phi}=1$ with effective equation of state parameter $\omega_{\phi}=\omega_{\rm eff}=1/3$. Its eigenvalues $(1, 2-\lambda/2)$ show that $A_{1}$ is an unstable node for $\lambda<4$ and a saddle for $\lambda>4$. Thus, it cannot act as a late time attractor, but it can act as early time repeller for $\lambda<4$ or a transient early time phase for $\lambda>4$.
    \item \textbf{Critical Point $A_{2}$:} Point $A_{2}$ also satisfies $\Omega_{\phi}=1$ and features an effective equation of state $\omega_{\rm eff}=\lambda/3-1$, covering de Sitter ($\lambda=0$), quintessence-like ($\lambda<2$), and phantom-like ($\lambda<0$) regimes. The eigenvalues $(\lambda-3)$ and $(\lambda-4)(\lambda-6)/(\lambda-8)$ imply stability for $\lambda<3$, and for $\lambda<2$ the point becomes a stable accelerating attractor. These properties make $A_{2}$ a strong candidate for the universe's late-time evolution.
    \item \textbf{Critical Point $A_{3}$:} The point $A_{3}$ yields $\Omega_{\phi}=1$ and a stiff-fluid equation of state $\omega_{\rm eff}=1$, indicating strong deceleration. Its eigenvalues $(3, \lambda-6)$ ensure at least one positive eigenvalue for all $\lambda$, rendering it a saddle for $\lambda<6$ and an unstable node for $\lambda>6$. Consequently, $A_{3}$ cannot serve as a late-time attractor and is relevant only as a possible early stage phase.
\end{itemize}
\begin{figure}[h]
    \centering
    \includegraphics[width=0.45\linewidth]{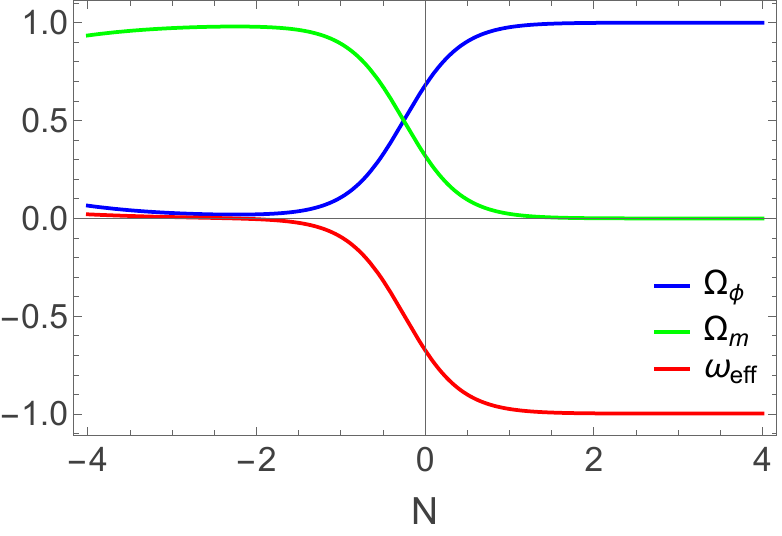}\hspace{0.5cm}
    \includegraphics[width=0.45\linewidth]{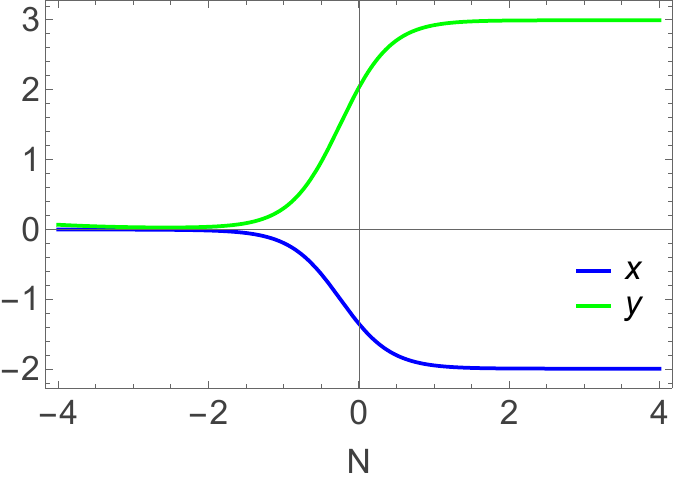}
    \caption{Evolution of $\Omega_\phi$, $\Omega_m$, $\omega_{\rm eff}$, $x$ and $y$ with the number of e-folds $N$.  
    The plots correspond to the initial conditions 
    $\Omega_{\phi}(0)=0.68$, $\omega_{\phi}(0)=-0.99$, and the parameter $\lambda=0.01$. This behavior mimics $P_1$ in the main text.}
    \label{fig:xy_evolution}
\end{figure}
Fig.\ref{fig:xy_evolution} illustrates the numerical evolution of the dynamical 
variables $\Omega_\phi$, $\Omega_m$, $\omega_{\rm eff}$, $x$, and $y$ with respect 
to the number of e-folds $N$. The plots correspond to the initial conditions 
$\Omega_\phi(0)=0.68$, $\omega_\phi(0)=-0.99$, and $\lambda = 0.01$. As shown in 
the left panel, the matter density parameter $\Omega_m$ decreases while 
$\Omega_\phi$ increases and asymptotically approaches unity, causing the effective 
equation of state $\omega_{\rm eff}$ to evolve towards $-1$, signalling the onset 
of a late-time accelerating phase. The right panel shows the evolution of the 
variables $x$ and $y$, which settle to constant values associated with the stable 
attractor $A_2$. Overall, the figure demonstrates the transition from a 
matter-dominated epoch to a scalar-field dominated accelerating regime for the 
chosen initial conditions.




\subsection{$V(\phi)=0$ and $\lambda$ as variable}
In this subsection we analyze the stability and cosmological interpretation of the fixed points of the
three–dimensional autonomous system in the variables $(x,y,\lambda)$ for the case $V=0$.
\begin{table}[h!]
\centering
\begin{tabular}{|c|c|c|c|c|c|}
\hline
\textbf{Points} & \textbf{$x$} & \textbf{$y$} & \textbf{$\lambda$} & \textbf{$\Omega_{\phi}$} & \textbf{$\omega_{\rm eff}$} \\
\hline
$B_{1}$ & $-2$ & $3$ & $0$ & $1$ & $-1$ \\
\hline
$B_{2}$ & $0$ & $1$ & $0$ & $1$ & $1/3$ \\
\hline
$B_{3}$ & $0$ & $1$ & $\dfrac{4}{3 - 4\rho}$ & $1$ & $1/3$ \\
\hline
$B_{4}$ & $1$ & $0$ & $0$ & $1$ & $1$ \\
\hline
\end{tabular}
\caption{Fixed points of the system and the corresponding value of $\Omega_{\phi}$, and $\omega_{\rm eff}$.}
\label{tab:FP_clean}
\end{table}
\begin{table}[h!]
\centering
\begin{tabular}{|c|c|c|}
\hline
\textbf{Points} & \textbf{Eigenvalues} & \textbf{Stability} \\
\hline
$B_{1}$ & $\left\{-3,\,-3,\,0\right\}$ &
Non hyperbolic \\
\hline
$B_{2}$ & $\left\{2,\,1,\,1\right\}$ &
Unstable \\
\hline
$B_{3}$ & $\left\{\dfrac{4-8\rho}{3-4\rho},\,-1,\,1\right\}$ &
Saddle \\
\hline
$B_{4}$ & $\left\{-6,\,3,\,0\right\}$ &
Non hyperbolic\\
\hline
\end{tabular}
\caption{Eigenvalues and stability condition of each fixed point.}
\label{tab:eigen_stability_rho}
\end{table}

\begin{itemize}
    \item \textbf{Critical Point \(B_1\):}
This is corresponding to a de Sitter phase as $\omega_{eff}=-1$. The eigenvalues $\{-3,-3,0\}$ indicate two stable directions and one marginal direction. Hence $B_1$ behaves as a late time attractor in the $(x,y)$ plane and is the only fixed point providing accelerated expansion. It is, therefore, the unique viable dark energy solution.
    \item \textbf{Critical Point \(B_2\):}
The point $B_2$ has all eigen values to be positive with effective equation of state parameter $\omega_{eff}=\frac{1}{3}$, which makes it highly unstable point, it can act as early time repeller.

 \item \textbf{Critical Point \(B_3\):}
This points has same behaviour as $P_2$. The eigenvalues $\left\{\frac{4-8\rho}{3-4\rho}, -1, 1 \right\}$ always include both positive and negative modes for all $\rho \neq \tfrac{3}{4}$; hence $B_3$ is a saddle point. It can act as a transient phase.

\item \textbf{Critical Point \(B_4\):}
This is corresponding to a stiff fluid $\omega_{eff}=1$ with mixed eigenvalues $\{-6,3,0\}$ classifying $B_4$ as a saddle point. It may appear only as a transient early epoch and does not describe the present Universe.
\end{itemize}

\begin{figure}[h]
    \centering
    \includegraphics[width=0.32\linewidth]{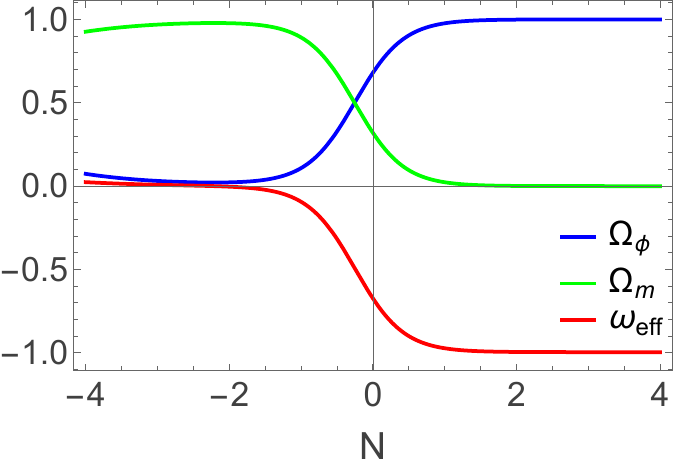}
    \includegraphics[width=0.31\linewidth]{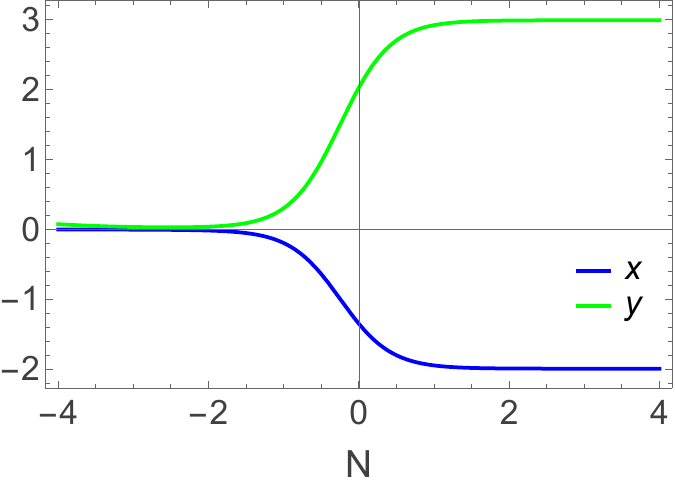}
    \includegraphics[width=0.34\linewidth]{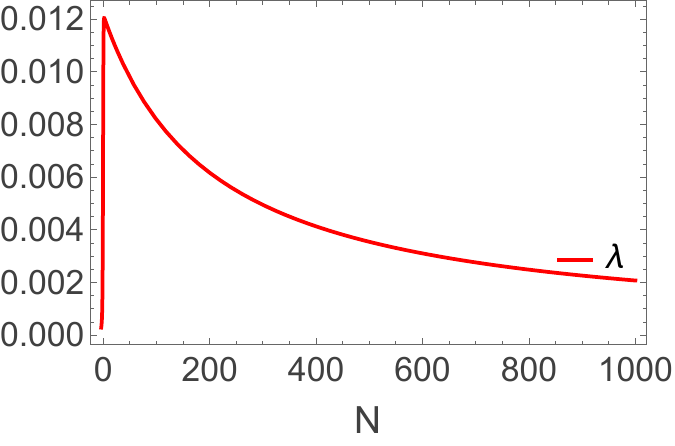}
    \caption{Evolution of $\Omega_\phi$, $\Omega_m$, $\omega_{\rm eff}$, $x$, $y$ and $\lambda$ with the number of e-folds $N$. 
    The plots correspond to the initial conditions 
    $\Omega_{\phi}(0)=0.68$, $\omega_{\phi}(0)=-0.99$, and the parameter $\rho=0.1$. This behavior mimics $P_1$ in the main text.}
    \label{fig:xylambda_evolution}
\end{figure}
Figure~\ref{fig:xylambda_evolution} shows how the quantities $\Omega_\phi$, $\Omega_m$, 
$\omega_{\rm eff}$, $x$, $y$ and $\lambda$ change with the number of e-folds $N$. The plots 
are generated using the initial conditions $\Omega_\phi(0)=0.68$, 
$\omega_\phi(0)=-0.99$, and $\rho = 0.1$. From the left panel, we see that the 
matter density $\Omega_m$ gradually decreases, while the scalar-field density 
$\Omega_\phi$ increases and approaches unity at late times. As this happens, the 
effective equation of state $\omega_{\rm eff}$ moves towards $-1$, indicating 
that the system enters an accelerated expansion phase. The right panel shows the 
evolution of the variables $x$ and $y$. Both of them smoothly settle to constant 
values, which correspond to the late-time attractor $B_1$. Overall, the figure 
shows that, for these initial conditions, the system naturally evolves from a 
matter-dominated stage to a scalar-field dominated accelerating phase.


\subsection{$V(\phi) \neq 0$ and $\lambda$ as Constant}
In this case we are considering x, y and z to be dynamical variables while keeping $\lambda$ to be a constant parameter. 
The critical points extracted from the set of equations (\ref{eq:d1}-\ref{eq:d3}) along with their respective eigenvalues are given in Table \ref{table:xyzpoints} and Table \ref{table:xyzeigenval}.

\begin{table}[H]
\centering
\begin{tabular}{|c|c|c|c|c|c|}
\hline
Points & $x$ & $y$ & $z$ & $\Omega_{\phi}$ & $\omega_{\text{eff}}$ \\
\hline
$C_{1}$ & $0$ & $1$ & $0$  & $1$ & $\frac{1}{3}$ \\[8pt]

$C_{2}$ & $0$ & $\frac{1}{4}(2\lambda - \sigma)$ & $\frac{1}{4}(4 - 2\lambda + \sigma)$ 
& $1$ & $\frac{1}{3} (2 \lambda -\sigma -3)$ \\[8pt]

$C_{3}$ & $\frac{1}{2}(-4 + \lambda)$ & $\frac{1}{2}(6 - \lambda)$ & $0$ 
& $1$ & $\frac{\lambda }{3}-1$ \\[12pt]

$C_{4}$ & $1$ & $0$ & $0$ 
& $1$ & $1$ \\[10pt]

$C_{5}$ & $\frac{1}{6}(2\lambda - \sigma)$ & $0$ & $\frac{1}{6}(6 - 2\lambda + \sigma)$ 
& $1$ & $\frac{2 \lambda }{3}-\frac{\sigma }{3}-1$ \\[12pt]
\hline
\end{tabular}
\caption{Critical points with corresponding $\Omega_{\phi}$ and $\omega_{\text{eff}}$.}
\label{table:xyzpoints}
\end{table}

\begin{table}[H]
\centering
\renewcommand{\arraystretch}{1.4}
\begin{tabular}{|c|c|c|}
\hline
Point & Eigenvalues & Stability \\ \hline

$C_1$ 
& $\{\,1,\;\tfrac{4-\lambda}{2},\;4-2\lambda+\sigma\,\}$ 
& Unstable \\ \hline

$C_2$ 
& $\{\,\tfrac{\lambda-\sigma}{2},\;-4+2\lambda-\sigma,\;-3+2\lambda-\sigma\,\}$ 
& Stable if $\sigma>\max\{\lambda,\,2\lambda-3\}$ \\ \hline

$C_3$ 
& $\{\,\lambda-3,\;\tfrac{(\lambda-4)(\lambda-6)}{\lambda-8},\;-\lambda+\sigma\,\}$ 
& Stable if $\lambda<3$ and $\sigma<\lambda$\\ \hline

$C_4$ 
& $\{\,3,\;-6+\lambda,\;6-2\lambda+\sigma\,\}$ 
& Unstable \\ \hline

$C_5$ 
& $\{\,-\lambda+\sigma,\;-6+2\lambda-\sigma,\;-3+2\lambda-\sigma\,\}$ 
& Stable if $\lambda>\sigma$ and $2\lambda-3<\sigma<\lambda$\\ \hline
\end{tabular}

\caption{Eigenvalues, and stability classification of the system depending on parameters $\lambda$ and $\sigma$.}
\label{table:xyzeigenval}
\end{table}


\begin{itemize}
    \item \textbf{Critical Point \( C_1 \):}  
The eigenvalues of \( C_1 \) are \( \left\{ 1,\; \frac{4-\lambda}{2},\; 4-2\lambda+\sigma \right\} \), where the positive eigenvalue \( 1 \) immediately indicates that the point is unstable/saddle. Additionally, at \( C_1 \), the effective equation of state \( w_{\text{eff}} = \frac{1}{3} \) corresponds to a decelerating phase. Therefore, this point is not suitable for modeling late-time acceleration. Instead, it can act as an early time repeller when \( \lambda < 4 \) and \( \sigma > 2\lambda - 4 \), while for other parameter choices it corresponds to a transient saddle phase.
    \item \textbf{Critical Point \( C_2 \):}  
The critical point $C_2$ has eigenvalues $\left\{ \frac{\lambda-\sigma}{2},\, -4+2\lambda-\sigma,\, -3+2\lambda-\sigma \right\}$, and stability requires all of them to be negative, which leads to the condition $\sigma>\max\{\lambda,\,2\lambda-3\}$. The effective equation of state is $w_{\text{eff}}=\frac{1}{3}(2\lambda-\sigma-3)$, and accelerated expansion requires $\sigma>2\lambda-2$. Therefore, critical point $C_2$ corresponds to a stable late-time accelerating solution for $\sigma>\max\{\lambda,\,2\lambda-2\}$.
    \item \textbf{Critical Point \( C_3 \):}  
The critical point $C_3$ has eigenvalues $\{\lambda-3,\; \frac{(\lambda-4)(\lambda-6)}{\lambda-8},\; -\lambda+\sigma\}$ and an effective equation of state $w_{\mathrm{eff}}=\frac{\lambda}{3}-1$. $C_3$ behaves as a stable late-time attractor if $\lambda<3$ and $\sigma>\lambda$ satisfied. In addition, the solution corresponds to an accelerated expansion phase when $\lambda<2$. Therefore, the critical point $C_3$ represents a late time stable accelerating solution for $\lambda<2$ and $\sigma>\lambda$.
    \item \textbf{Critical Point \( C_4 \):}  
The eigenvalues of \( C_4 \) are \( \{3, \, -6 + \lambda, \, 6 - 2\lambda + \sigma \} \), where the positive eigenvalue \( 3 \) immediately indicates that the point is unstable/saddle. Additionally, at \( C_4 \), the effective equation of state \( w_{\text{eff}} = 1 \) corresponds to a decelerating phase, incompatible with the late-time accelerated expansion seen in the universe. Therefore, this point is not suitable for modeling late-time acceleration. Instead, it can act as an early time repeller when \( \lambda > 6 \) and \( \sigma < 2\lambda -6 \), while for other parameter choices it corresponds to a transient saddle phase.
    \item \textbf{Critical Point \( C_5 \):}  
The critical point $C_5$ is defined by the eigenvalues $(-\lambda+\sigma)$, $(-6+2\lambda-\sigma)$, and $(-3+2\lambda-\sigma)$, with an effective equation of state given by $w_{\mathrm{eff}}=-1+\frac{2\lambda-\sigma}{3}$. This point is stable when all eigenvalues are negative, which occurs for $2\lambda-3<\sigma<\lambda$, implying that the system naturally evolves toward $C_5$ at late times. Moreover, the solution corresponds to an accelerating universe when $w_{\mathrm{eff}}<-1/3$, or equivalently when $\sigma>2\lambda-2$. As a result, for parameter values satisfying $2\lambda-2<\sigma<\lambda$, the critical point describes the late time accelerated phase.
\end{itemize}

\begin{figure}[ht!]
    \centering

    \begin{subfigure}{0.31\linewidth}
        \centering
        \includegraphics[width=\linewidth]{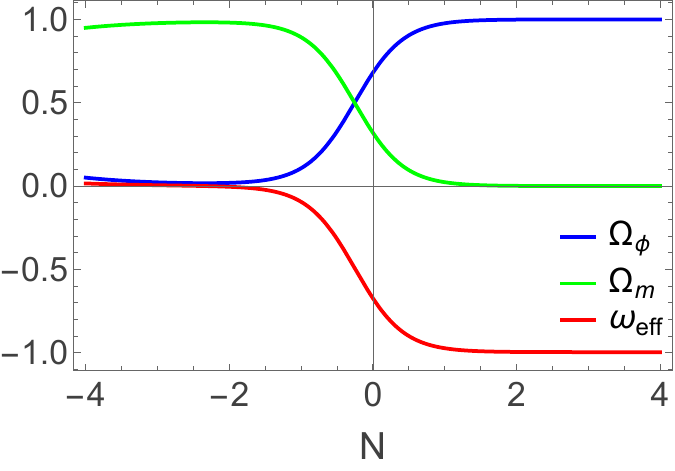}
        \caption{}
    \end{subfigure}
    \hspace{0.25cm}
    \begin{subfigure}{0.3\linewidth}
        \centering
        \includegraphics[width=\linewidth]{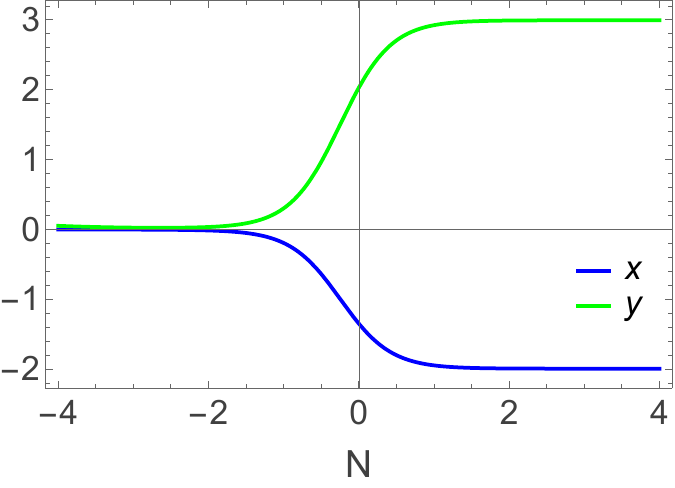}
        \caption{}
    \end{subfigure}
    \hspace{0.25cm}
    \begin{subfigure}{0.34\linewidth}
        \centering
        \includegraphics[width=\linewidth]{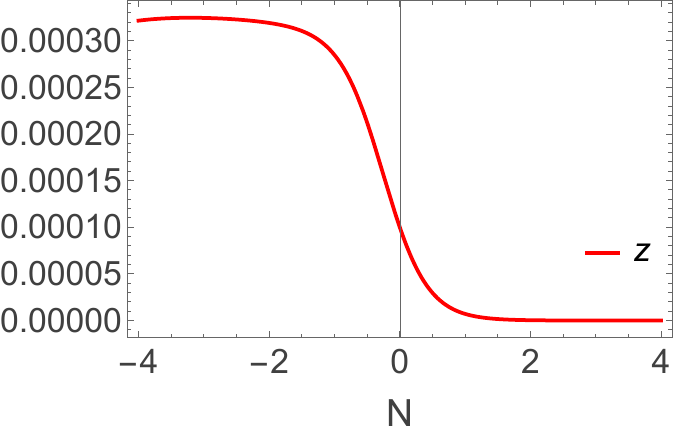}
        \caption{}
    \end{subfigure}

    \caption{Evolution of $\Omega_\phi$, $\Omega_m$, $\omega_{\rm eff}$, $x$, $y$, and $z$ with the number of e-folds $N$,
    for initial conditions 
    $\Omega_{\phi}(0)=0.68$, $\omega_{\phi}(0)=-0.99$, 
    $z(0)=10^{-4}$, the parameters $\sigma=-3$, and $\lambda=0.01$. This behavior also mimics $P_1$ in the main text.}
    \label{fig:xyz_evolution}
\end{figure}
Figure \ref{fig:xyz_evolution} shows how the system evolves for the chosen parameters $\lambda = 0.01$ and $\sigma = -3$. The plots clearly indicate that the universe moves from an initial transient stage towards a stable final state. The scalar-field density $\Omega_{\phi}$ slowly increases and becomes dominant, while the matter density $\Omega_{m}$ decreases, and the effective equation of state approaches a value very close to $-1$, showing that accelerated expansion is reached. The variables $x$ and $y$ also settle to constant values after some time, which means that the system is approaching a fixed point. The variable $z$ quickly decreases to zero, matching the prediction from the stability analysis. Taken together, all three figures confirm that the system evolves towards the only stable critical point for these parameter values, $C_{3}$, which represents a late-time accelerating solution.


\begin{figure}[ht!]
    \centering

    \begin{subfigure}{0.31\linewidth}
        \centering
        \includegraphics[width=\linewidth]{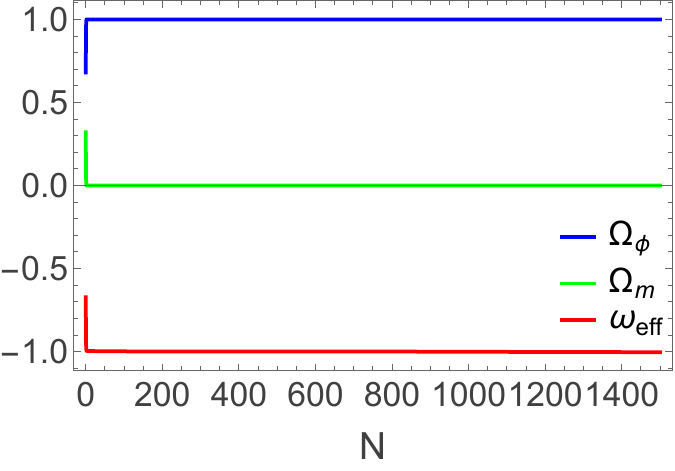}
        \caption{}
        \label{fig:xyzP}
    \end{subfigure}
    \hspace{0.25cm}
    \begin{subfigure}{0.31\linewidth}
        \centering
        \includegraphics[width=\linewidth]{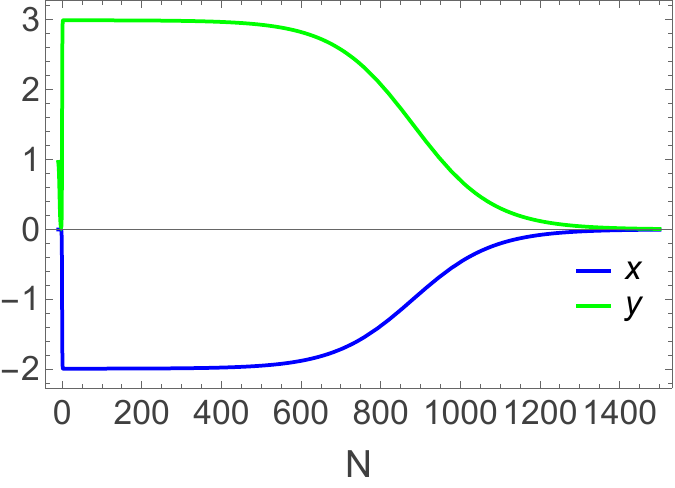}
        \caption{}
        \label{fig:xyzQ}
    \end{subfigure}
    \hspace{0.25cm}
    \begin{subfigure}{0.31\linewidth}
        \centering
        \includegraphics[width=\linewidth]{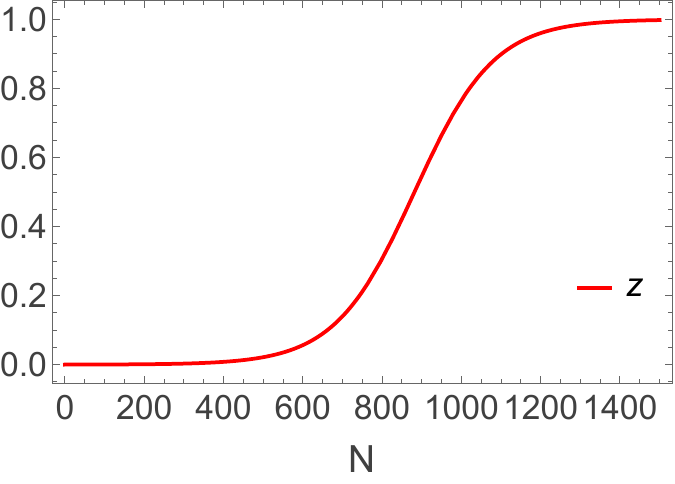}
        \caption{}
        \label{fig:xyzR}
    \end{subfigure}

    \caption{Evolution of $\Omega_\phi$, $\Omega_m$, $\omega_{\rm eff}$, $x$, $y$, and $z$
    with the number of e-folds $N$, for initial conditions $\Omega_{\phi}(0)=0.68$, 
    $\omega_{\phi}(0)=-0.99$, 
    $z(0)=10^{-4}$, the parameters 
    $\sigma = 0.02$, 
    and $\lambda = 0.01$. This behavior mimics $P_3$ in the main text.}
    \label{fig:xyz_subfigures}
\end{figure}
Figure \ref{fig:xyz_subfigures} illustrates the dynamical evolution of the system for the initial conditions $\Omega_{\phi}(0)=0.68$, $\omega_{\phi}(0)=-0.99$ and $z(0)=10^{-4}$, with the parameter values $\sigma=0.02$ and $\lambda=0.01$. Figure 4(a) shows that the scalar-field energy density $\Omega_{\phi}$ approaches unity, while the matter density $\Omega_{m}$ decays to zero, causing the effective equation of state to settle at $\omega_{\rm eff}=-1$, which is indicative of a accelerated expansion. Figure 4(b) displays the evolution of the dynamical variables $x$ and $y$, both of which approach towards zero as the system goes in far future. Figure 4(c) shows that the variable $z$ grows monotonically from its initially negligible value and asymptotically approaches $z\simeq 1$. Altogether, these behaviors confirm that the system evolves toward the stable late-time attractor $C_{2}$, corresponding to a scalar-field dominated Universe with effective cosmological-constant behavior.

\bibliographystyle{unsrtnat}
\bibliography{refs}

\end{document}